\documentclass[sigconf]{acmart}

\usepackage{booktabs}
\usepackage{amsmath}
\usepackage{multirow}

\title{Rethinking Item Tokenization in Generative Recommenders: From Fixed Atoms to Semantic Subwords}

\copyrightyear{2026}
\acmYear{2026}
\setcopyright{cc}
\setcctype{by}
\acmConference[CIKM '26] {Proceedings of the 35th ACM International Conference on Information and Knowledge Management}{November 7--11, 2026}{Rome, Italy.}
\acmBooktitle{Proceedings of the 35th ACM International Conference on Information and Knowledge Management (CIKM '26), November 7--11, 2026, Rome, Italy}
\acmISBN{979-8-4007-2539-5/2026/11}
\acmDOI{10.1145/3799682.3840979}
\author{Xinrui Miao}
\orcid{0009-0003-8848-7457}
\affiliation{
  \institution{University of Science and Technology of China}
  \city{Hefei}
  \country{China}
}
\email{miaoxinrui@mail.ustc.edu.cn}

\author{Mingjia Yin}
\orcid{0009-0005-0853-1089}
\affiliation{
  \institution{University of Science and Technology of China}
  \city{Hefei}
  \country{China}
}
\email{mingjia-yin@mail.ustc.edu.cn}

\author{Jiaqing Zhang}
\orcid{0009-0001-1039-9735}
\affiliation{
  \institution{University of Science and Technology of China}
  \city{Hefei}
  \country{China}
}
\email{jiaqing.zhang@mail.ustc.edu.cn}

\author{Wei Guo}
\orcid{0000-0001-8616-0221}
\affiliation{
  \institution{Huawei Technologies}
  \city{Shanghai}
  \country{China}
}
\email{guowei67@huawei.com}

\author{Yong Liu}
\orcid{0000-0001-9031-9696}
\affiliation{
  \institution{Huawei Technologies}
  \city{Shanghai}
  \country{China}
}
\email{liu.yong6@huawei.com}

\author{Yuyang Ye}
\orcid{0000-0002-1513-7814}
\affiliation{
  \institution{University of Science and Technology of China}
  \city{Hefei}
  \country{China}
}
\email{yeyuyang@mail.ustc.edu.cn}

\author{Hao Wang}
\correspondingauthor
\orcid{0000-0001-9921-2078}
\affiliation{
  \institution{University of Science and Technology of China}
  \city{Hefei}
  \country{China}
}
\email{wanghao3@ustc.edu.cn}

\author{Enhong Chen}
\orcid{0000-0002-4835-4102}
\affiliation{
  \institution{University of Science and Technology of China}
  \city{Hefei}
  \country{China}
}
\email{cheneh@ustc.edu.cn}

\renewcommand{\shortauthors}{Miao et al.}

\begin{abstract}
In generative recommender systems, items are typically tokenized into fixed-length semantic ID sequences for autoregressive next-item prediction. However, for user-context modeling, this fine-grained representation triggers \textbf{Intra-item Attention Overload}: excessive attention is spent on low-level intra-item dependencies rather than high-level inter-item behavioral transitions.

To address this, we propose \textbf{Semantic Subword Tokenization (SST)}, which represents historical items as variable-length semantic subwords while preserving fixed-length target decoding. SST first applies \textbf{Item-level Subword Tokenization (IST)} to merge stable adjacent \textbf{atom tokens} into compact \textbf{semantic subword tokens}, thereby reducing intra-item reassembly in the encoder. It then introduces \textbf{Behavior-induced Co-occurrence Augmentation (BCA)} to inject coarse-grained semantic prefix transition signals, guiding the freed modeling capacity toward inter-item behavioral regularities. Extensive experiments on three public datasets and three generative recommender backbones show empirical improvements of SST over fixed-length and transferable variable-length SID baselines. Code is available at \url{https://github.com/mxrcandy/Semantic-Subword-Tokenization}.
\end{abstract}



\begin{document}

\maketitle

\section{Introduction}
\label{sec:intro}
\sloppy

Generative recommendation has emerged as a prominent paradigm for next-item recommendation, with Semantic ID (SID)-based methods forming a widely adopted stream~\cite{rajput2023recommender,zheng2024adapting,wang2024learnable,hou2025survey}. This paradigm is driven by advances in large language models (LLMs)~\cite{zhao2023survey,floridi2020gpt,achiam2023gpt} and generative retrieval~\cite{tay2022transformer,wang2022neural,sun2023learning}. Instead of ranking items by nearest-neighbor retrieval over single item IDs~\cite{kang2018self,hidasi2015session,sun2019bert4rec,tang2018personalized}, these methods represent each item as a compact sequence of discrete semantic tokens and train a generative model to produce the SID of the next item. In particular, residual-quantization-based tokenizers map item representations into fixed-length codebook sequences, such as $\langle a\rangle\langle b\rangle\langle c\rangle\langle d\rangle$. We refer to each individual codebook token as an \textbf{atom token}. Fixed-length SIDs have become a common design choice because they provide every item with a uniform decoding grammar, making autoregressive generation and constrained beam search straightforward.

\begin{figure}[t]
\centering
\includegraphics[width=\columnwidth]{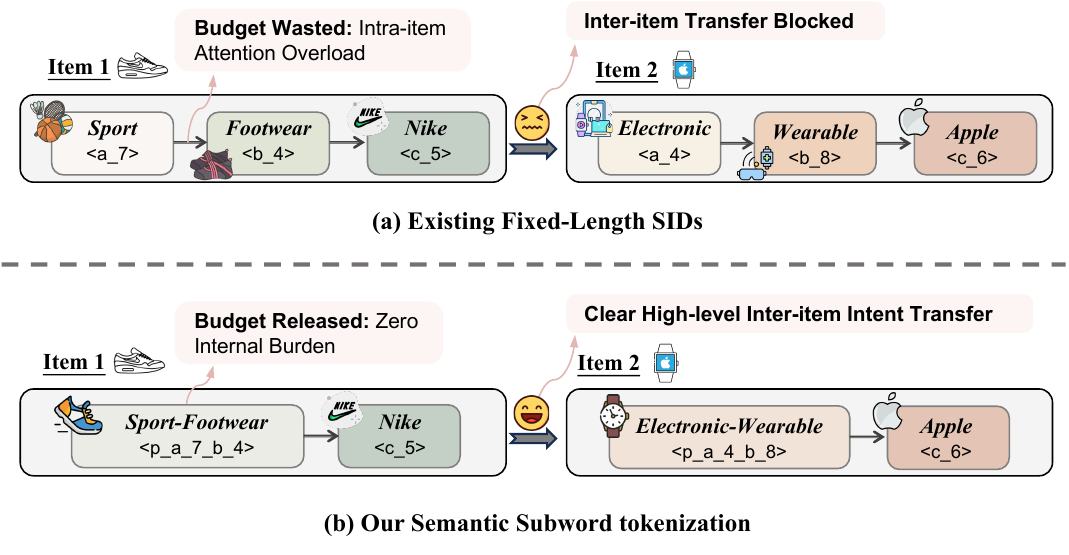}
\caption{Motivation of our work. Fixed-length SIDs are useful target-side decoding grammars, but when reused to represent user context they expose low-level atom-token dependencies to the encoder. SST decouples these two roles by retokenizing historical items into semantic subwords while preserving fixed-length target decoding.}
\Description{A two-part comparison of two historical items. The upper part uses four atom tokens per item and labels the area between items as attention budget wasted on intra-item overload, which blocks high-level inter-item transfer. The lower part merges adjacent atom pairs into semantic subword tokens, labels the internal burden as zero, and shows a clear high-level intent transfer between the two items.}
\label{fig:motivation}
\end{figure}

However, a decoding-oriented identifier is not necessarily an effective representation for modeling user context. In standard SID-based generative recommenders, the same fixed-length atom sequence used to identify the target item is also reused to represent each historical item in the user context. After user histories are flattened into token sequences, every historical item is decomposed into multiple atom tokens. This exposes short-range, intra-item dependencies directly to the sequence encoder: before the model can reason about transitions between items, it must repeatedly compose atom tokens that belong to the same item into an item-level semantic signal. We call this user-context tokenization bottleneck \textbf{Intra-item Attention Overload}: fixed-length SIDs impose an item-reassembly workload on the encoder, causing a nontrivial portion of its attention and modeling capacity to be spent on low-level intra-item token relations rather than higher-level behavioral transitions. Figure~\ref{fig:motivation} illustrates this mismatch between fixed-length SIDs as target identifiers and fixed-length SIDs as representations for modeling user context.

This bottleneck is especially problematic because user preference modeling is fundamentally about inter-item dynamics. For example, a coarse transition from sports equipment to wearable electronics should be represented as a high-level behavioral relation, but under fixed-length SIDs it is observed by the model as several atom-level token transitions. The resulting sequence mixes item-internal reconstruction signals with inter-item intent signals. In Section~\ref{sec:attention-overload}, we empirically examine this phenomenon through a token-level attention workload analysis, showing that fixed-length SIDs allocate substantial intra-item attention mass to atom-level item reassembly within user sequences.

Existing efforts improve generative recommenders from adjacent but distinct perspectives. Text-based identifiers replace discrete item codes with natural-language tokens~\cite{geng2022recommendation,cui2022m6,tan2024idgenrec}, but often lead to long input sequences and less stable target grounding. SID optimization methods improve codebook utilization, semantic alignment, collision handling, or tokenizer--recommender consistency~\cite{wang2024learnable,zhou2025onerec,hu2026stop,liu2025generative}; however, they still represent historical items using the same fine-grained atom sequences designed for target generation. Consequently, the attention workload caused by atom-level item reassembly remains largely unaddressed.

To address this problem, we propose \textbf{Semantic Subword Tokenization (SST)}, which retokenizes historical items into variable-length semantic subwords for modeling user context. SST first applies \textbf{Item-level Subword Tokenization (IST)}, which learns merge rules over stable adjacent spans of atom tokens and rewrites historical item SIDs into compact \textbf{semantic subword tokens}. By encoding frequent intra-item token couplings directly into the input tokenization, IST reduces the need for the encoder to repeatedly perform atom-level item reassembly and shortens the user-context token sequence. The target item is still generated with the original fixed-length SID.

IST removes part of the low-level item-reassembly burden, but compression alone does not tell the model which inter-item regularities are important for recommendation. This motivates \textbf{Behavior-induced Co-occurrence Augmentation (BCA)}, a behavior-guided stage that mines coarse-grained \textbf{semantic prefix transitions} from user behavior sequences and injects the corresponding replay samples into training. These prefix-level co-occurrences expose recurring intent transitions in user behavior, complementing the intra-item compression introduced by IST.

A design principle of SST is asymmetric tokenization. Since the user context is consumed by the encoder as conditioning information, semantic subword retokenization reduces the item-reassembly workload in context modeling. The target side instead defines the autoregressive decoding grammar for item identification, so we retain the fixed-length SID for stable target generation.

In summary, our main contributions are as follows:
\begin{itemize}
\item We identify a user-context tokenization bottleneck in fixed-length SID-based generative recommendation: decoding-oriented atom sequences are reused to represent historical items, causing intra-item attention overload and distracting the encoder from inter-item behavioral transitions.

\item We propose Semantic Subword Tokenization (SST), which retokenizes historical items into semantic subword tokens via IST while preserving fixed-length target decoding. BCA
further supplies behavior-level semantic prefix transition signals that complement the intra-item compression introduced by IST. Through these two components, SST reduces item-reassembly workload and redirects modeling capacity toward higher-order user intent.

\item We evaluate SST on three public datasets and three mainstream backbones, showing consistent recommendation gains, reduced intra-item attention workload, and the effectiveness of the asymmetric history-side tokenization design in the evaluated settings.
\end{itemize}

\section{Methodology}
\label{sec:method}
\sloppy

\begin{figure*}[t]
\centering
\includegraphics[width=\textwidth]{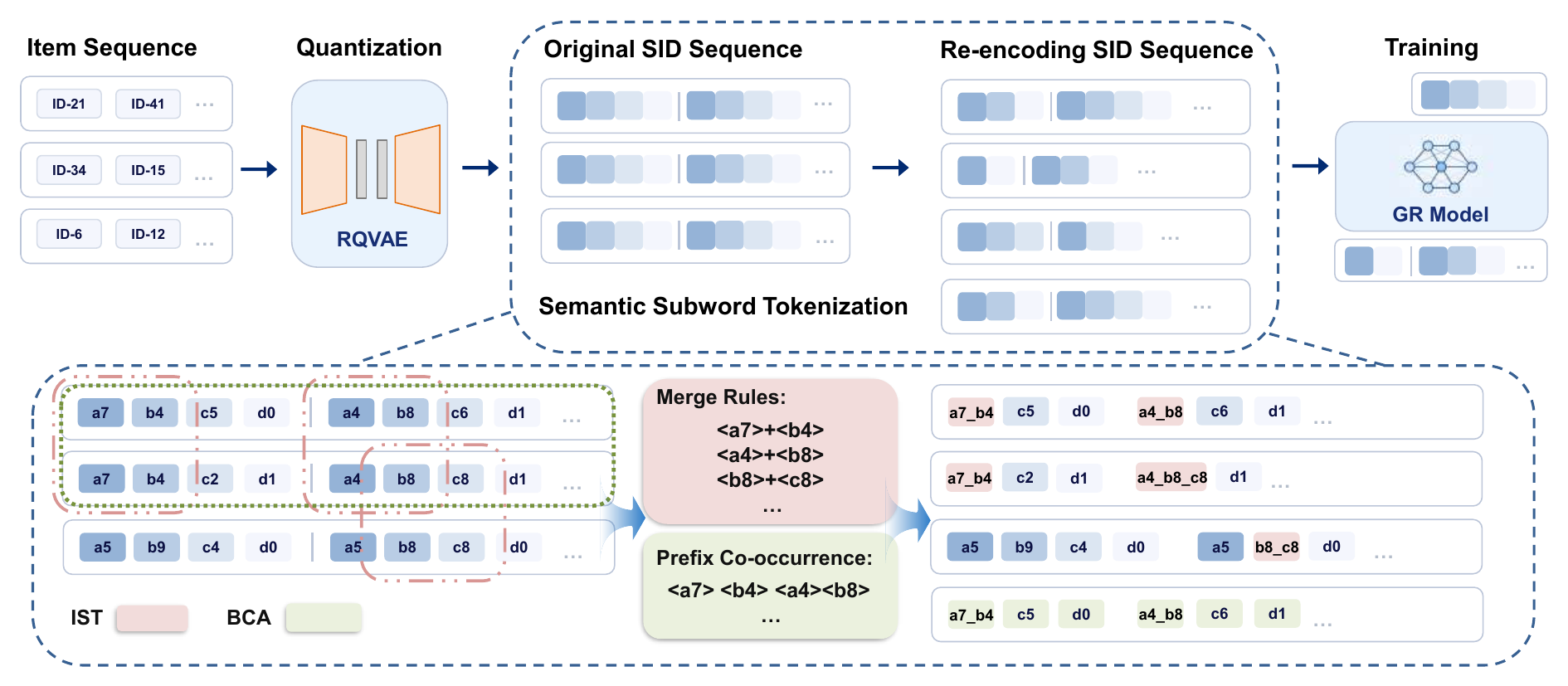}
\caption{Overview of the proposed Semantic Subword Tokenization (SST) framework. IST learns item-level semantic subword tokens by merging stable adjacent SID tokens, while BCA injects coarse-grained behavior co-occurrence signals to guide the attention capacity released by IST.}
\Description{A workflow diagram for SST. Fixed-length semantic IDs are first processed by Item-level Subword Tokenization, which learns adjacent-token merge rules and rewrites history items as variable-length semantic subwords. Behavior-induced Co-occurrence Augmentation then extracts semantic prefixes, mines co-occurring prefix pairs, and replays their behavior subsequences during training. The history encoder receives subword tokens, while the target decoder continues to generate fixed-length semantic IDs.}
\label{fig:sst-method-overview}
\end{figure*}

This section first formulates the generative recommendation task and reviews fixed-length SIDs from the perspective of user-context tokenization, then presents our Semantic Subword Tokenization (SST) method, including Item-level Subword Tokenization (IST) and Behavior-induced Co-occurrence Augmentation (BCA). Figure~\ref{fig:sst-method-overview} provides an overview of the proposed framework.

\subsection{Problem Formulation}
\label{sec:problem}

Let the item universe be $\mathcal{V} = \{v_1, v_2, \ldots, v_{|\mathcal{V}|}\}$. For each user $u$, their interaction history sorted by timestamp forms a sequence $S_u = [v_1^{(u)}, v_2^{(u)}, \ldots, v_{n_u}^{(u)}]$. Each item is represented by a discrete token sequence. The goal of generative recommendation is to model the conditional probability distribution $P(v_{n_u+1}^{(u)} \mid S_u)$ and autoregressively generate the semantic ID sequence of the next item for future interactions token by token~\cite{hou2025survey}.

\subsection{Preliminaries: Fixed-Length SIDs and User-Context Tokenization}
\label{sec:prelim}

We review the fixed-length SID pipeline used by mainstream generative recommenders, focusing on how item identifiers are produced and how they are reused to represent user histories.

\subsubsection{Fixed-Length SID Tokenization}

The item tokenizer quantizes item semantic features (typically dense embeddings produced by a pre-trained large language model) into discrete semantic ID sequences~\cite{li2025survey}. Mainstream methods adopt the \textbf{Residual Quantization (RQ)} framework~\cite{lee2022autoregressive, van2017neural,zeghidour2021soundstream}, which uses $L$ codebook layers to iteratively approximate an item embedding $\mathbf{e}_i$~\cite{ju2025generative}. At layer $j$, the tokenizer selects the nearest codeword to the current residual:

\begin{equation}
\mathbf{r}_0 = \mathbf{e}_i, \quad
k_j = \arg\min_{k \in \mathcal{C}_j} \|\mathbf{r}_{j-1} - \mathbf{c}_j^k\|_2^2, \quad
\mathbf{r}_j = \mathbf{r}_{j-1} - \mathbf{c}_j^{k_j}.
\label{eq:rq}
\end{equation}

The final item identifier is $\mathbf{t}_i = [t_i^1, t_i^2, \ldots, t_i^L]$, where each $t_i^j$ is an \textbf{atom token} from the $j$-th codebook. These atom tokens jointly identify the item in a coarse-to-fine manner: earlier tokens capture broader semantic regions, while later tokens refine residual details needed to distinguish individual items. In practice, the codebooks can be learned by RQ-VAE or RQ-KMeans; SST is agnostic to this tokenizer choice by design.

\subsubsection{Flattened User-Context Tokenization}

In SID-based generative recommenders, the same fixed-length SID serves two roles. On the target side, it defines a stable $L$-slot decoding grammar: the decoder autoregressively generates $L$ atom tokens, which are mapped back to an item through constrained decoding, typically with beam search~\cite{lowerre1976harpy}. On the history side, each interacted item is also represented by its full atom-token sequence. Thus a user history $S_u = [v_1^{(u)}, \ldots, v_{n_u}^{(u)}]$ is flattened into
$T_u = [t_1^{1}, \ldots, t_1^{L}, \ldots, t_{n_u}^{1}, \ldots, t_{n_u}^{L}]$.
Let $\mathbf{t}_{n_u+1}=[t_{n_u+1}^{1},\ldots,t_{n_u+1}^{L}]$ denote the fixed-length SID of the next item. Conditioned on the encoded history $T_u$, an encoder--decoder Transformer autoregressively predicts this target SID using the target-side next-token cross-entropy loss~\cite{hou2025survey}:

\begin{equation}
\mathcal{L}_{\text{GR}}
= -\sum_{j=1}^{L} \log P_\theta\!\left(
t_{n_u+1}^{j} \mid T_u, t_{n_u+1}^{<j}
\right)
\label{eq:loss}
\end{equation}

This flattening increases the context length from $n_u$ items to $n_uL$ tokens and exposes intra-item atom dependencies to the encoder. Before modeling inter-item behavioral transitions, the model must repeatedly spend attention reassembling atom tokens belonging to the same item into an item-level semantic signal. This creates the user-context tokenization bottleneck studied in this paper: fixed-length SIDs are convenient target identifiers, but inefficient history-side representations for modeling user behavior.

\subsection{Item-level Subword Tokenization (IST)}
\label{sec:ist}

To alleviate the user-context tokenization bottleneck, IST augments fixed-length SIDs with a post-hoc variable-length encoding stage. Inspired by subword tokenization in large language models~\cite{rajaraman2024toward,goldman2024unpacking,erdogan2026information}, IST packs frequently co-occurring contiguous atom-token spans into \textbf{semantic subword tokens}, so that low-level compositional regularities are encoded directly in the token structure rather than repeatedly relearned by the backbone.
Given the fixed-length SID of each item, IST learns merge rules and constructs a variable-length token vocabulary in three stages.

\noindent\textbf{Step 1: Candidate pair collection.}
For each item, we collect adjacent token pairs such as $(\langle a\_7\rangle,\langle b\_4\rangle)$; pairs crossing item boundaries are never considered. We then record two corpus-level statistics for each pair: its occurrence frequency and its item support, namely the number of distinct items containing it.

\noindent\textbf{Step 2: Merge rule learning.}
IST iteratively learns merge rules according to the chosen tokenization criterion. At each iteration, it selects the highest-scoring adjacent candidate whose frequency and item support satisfy the thresholds $f$ and $s$, respectively, merges it in all item SIDs, and updates the pair statistics. The procedure stops when the configured merge budget is reached or no eligible candidate remains. In practice, we consider three criteria: two adapted from LLM subword tokenization, BPE~\cite{sennrich2016neural} and WordPiece~\cite{devlin2019bert}, and one criterion designed for generative recommendation, namely the proposed CondEntropy.

\begin{itemize}
\item \textbf{BPE}: Following the subword tokenization paradigm used in NLP/LLMs, BPE repeatedly selects the most frequent adjacent token pair. After each selected pair is merged in all item SIDs, the pair statistics are updated before the next merge rule is learned. For a candidate pair $(t^a, t^b)$:
\begin{equation}
s_{\text{BPE}}(t^a, t^b) = \text{freq}(t^a, t^b)
\label{eq:bpe}
\end{equation}
where $\text{freq}(t^a, t^b)$ is the co-occurrence count of the adjacent pair in the item corpus. BPE is the simplest and most data-driven criterion, but it tends to favor token pairs involving already high-frequency tokens, which may dilute the semantic specificity of the resulting subwords.

\item \textbf{WordPiece}: WordPiece follows the same iterative merge process, but scores candidate pairs by PMI-weighted frequency. This penalizes generic high-frequency tokens and prioritizes pairs with stronger semantic coupling:
\begin{equation}
\begin{aligned}
s_{\text{WP}}(t^a, t^b)
&= \text{freq}(t^a, t^b)\cdot \max\!\bigl(\text{PMI}(t^a, t^b), 0\bigr),\\
\text{PMI}(t^a, t^b)
&= \log \frac{(\text{freq}(t^a, t^b)+1)N}{(\text{freq}(t^a)+1)(\text{freq}(t^b)+1)}.
\end{aligned}
\label{eq:wordpiece}
\end{equation}
where $N$ is the total number of adjacent token pairs. By combining frequency with positive PMI, WordPiece suppresses generic high-frequency pairs and favors discriminative associations among tokens.

\item \textbf{CondEntropy}: CondEntropy is designed for generative recommendation, where a useful subword should be both frequent and stable under autoregressive generation. Given the pair frequency $\text{freq}(t^a,t^b)$ defined above, we compute its next-side and previous-side marginal frequencies as $\text{freq}_{\text{next}}(t^a)=\sum_{z\in\mathcal{V}_t}\text{freq}(t^a,z)$ and $\text{freq}_{\text{prev}}(t^b)=\sum_{z\in\mathcal{V}_t}\text{freq}(z,t^b)$, where $\mathcal{V}_t$ denotes the current token vocabulary. The bidirectional empirical probabilities are
\begin{equation}
P(t^b\!\mid\!t^a)=\frac{\text{freq}(t^a,t^b)}{\text{freq}_{\text{next}}(t^a)}, \quad
P(t^a\!\mid\!t^b)=\frac{\text{freq}(t^a,t^b)}{\text{freq}_{\text{prev}}(t^b)}.
\label{eq:condentropy-prob}
\end{equation}
We use $\pi_{\text{next}}(t^a)$ and $\pi_{\text{prev}}(t^b)$ to denote the purity of the next-token distribution of $t^a$ and the previous-token distribution of $t^b$, respectively. Each purity term is computed as one minus the normalized conditional entropy over observed neighbors; a token with only one observed neighbor receives purity one. CondEntropy then scores each pair as
\begin{equation}
\begin{aligned}
s_{\text{CE}}(t^a, t^b)
&= \text{freq}(t^a, t^b)
   \cdot \sqrt{P(t^b\!\mid\!t^a)\,P(t^a\!\mid\!t^b)}\\
&\quad \cdot \sqrt{\pi_{\text{next}}(t^a)\,\pi_{\text{prev}}(t^b)}.
\end{aligned}
\label{eq:condentropy}
\end{equation}
Compared with frequency-only or PMI-based scoring, CondEntropy imposes a stricter quality bar: a pair scores high only when it is frequent, mutually predictive, and located in low-entropy local contexts, making the selected semantic subword more reliable.
\end{itemize}

\noindent\textbf{Step 3: Re-encoding.}
Finally, each item SID is rewritten by replaying the learned merge rules in order. Whenever a selected adjacent pair is merged, the resulting segment is assigned a unique \textbf{semantic subword token} (e.g., $\langle p\_a7\_b4\rangle$), which is added to the token vocabulary and trained together with atom tokens in the recommender. Tokens that are not covered by any merge rule remain unchanged. After re-encoding, the item sequence length is no longer fixed at $L$ but becomes variable, denoted by $\tilde{L}_i \leq L$.

\noindent\textbf{\textit{Remark.}} By consolidating short-range token co-occurrence regularities into semantic subword tokens, IST reduces the need for the Transformer to repeatedly fit low-level atom couplings in the attention layers. The resulting history context is more compact and semantically organized, yielding a concise representation for modeling inter-item transitions and higher-order user intent.

\subsection{Behavior-induced Co-occurrence Augmentation (BCA)}
\label{sec:bca}

IST reduces the low-level attention burden by compacting stable intra-item token co-occurrences into semantic subwords, but it does not specify which inter-item behavioral regularities should be emphasized. Therefore, BCA builds upon this cognitive offloading effect by injecting coarse-grained semantic prefix co-occurrence signals into the training process, providing structured high-order navigation signals for the model.
The procedure consists of the following steps.

\noindent\textbf{Step 1: Semantic prefix extraction.}
For each item in a user sequence, the first $P$ tokens of its SID are extracted as the semantic prefix, representing the item's coarse-grained semantic category. For example, given an item SID $\langle a\_7\rangle\langle b\_4\rangle\langle c\_5\rangle\langle d\_0\rangle$, extracting $P=2$ yields the prefix $\langle a\_7\rangle\langle b\_4\rangle$, which denotes its high-level semantic class.

\noindent\textbf{Step 2: Prefix pair co-occurrence mining.}
On each user's prefix sequence, a sliding window of size $W$ counts co-occurring prefix pairs $(p_{\text{left}}, p_{\text{right}})$. For each pair, we record its raw count, Pointwise Mutual Information (PMI), and lift, where PMI measures log association relative to the two prefix marginals and lift is its exponentiated ratio. The top-$k$ prefix pairs are selected by a specified scoring function (e.g., count, PMI-weighted, or lift). We adopt pairwise prefix co-occurrence as a simple and efficient behavioral signal: it is computationally lightweight and statistically reliable under sparse interaction data.

\noindent\textbf{Step 3: Sequence replay augmentation.}
For each selected high-frequency prefix pair $(p_{\text{left}}, p_{\text{right}})$, we trace back the context in which this transition occurred in the original user sequences---i.e., we extract the subsequence of historical items between the occurrence of $p_{\text{left}}$ and $p_{\text{right}}$. These subsequences are injected into the original training set as additional training samples. Each augmented sample uses the replayed historical subsequence as input and the right-side item as the prediction target.

\noindent\textbf{\textit{Remark.}} IST and BCA follow a clear progression: IST offloads low-level intra-item regularities, making room for the model to better absorb the coarse-grained co-occurrence signals injected by BCA; conversely, BCA additionally exposes the model to inter-item behavioral regularities. Together they form a complete ``remove low-order---inject high-order'' representation reinforcement loop for user modeling.

\subsection{Asymmetric Tokenization}
It is worth emphasizing that IST and BCA only modify the token representations on the encoder (history) side; the decoder (target) side consistently retains the fixed-length SID format. This asymmetric tokenization design is motivated by the following observation: while the history side benefits from context compression and structured representation, converting the target side to variable-length subword representations would disrupt the stable L-slot decoding grammar and force the decoder into a low-frequency competition between atom tokens and semantic subword tokens, ultimately harming generation stability. The necessity of retaining fixed-length SIDs on the target side is empirically validated and analyzed in Section~\ref{sec:asymmetric-tokenization}.

\section{Experiments}
\label{sec:experiments}
\sloppy

\begin{table*}[t!]
\fontsize{8.5pt}{9.8pt}\selectfont
\centering
\caption{Main experimental results. N@5/N@10 abbreviate NDCG@5/NDCG@10. The ``Improve.'' row denotes the relative improvement of SST (+IST+BCA) over the strongest non-SST baseline in each column.}
\label{tab:main-results}
\setlength{\tabcolsep}{3.4pt}
\renewcommand{\arraystretch}{1.08}
\begin{tabular}{ll|rrrr|rrrr|rrrr}
\toprule
& & \multicolumn{4}{c|}{Beauty} & \multicolumn{4}{c|}{Instruments} & \multicolumn{4}{c}{Yelp} \\
\cmidrule(lr){3-6} \cmidrule(lr){7-10} \cmidrule(lr){11-14}
Method & & HR@5 & HR@10 & N@5 & N@10 & HR@5 & HR@10 & N@5 & N@10 & HR@5 & HR@10 & N@5 & N@10 \\
\midrule
\multirow{7}{*}{TIGER-KM}
& Fixed    & 0.0372 & 0.0582 & 0.0238 & 0.0305 & 0.0849 & 0.1065 & 0.0718 & 0.0787 & 0.0244 & 0.0385 & 0.0159 & 0.0205 \\
& SARQ    & 0.0368 & 0.0579 & 0.0234 & 0.0302 & 0.0850 & 0.1048 & 0.0630 & 0.0694 & 0.0245 & 0.0390 & 0.0159 & 0.0206 \\
& VSID  & 0.0383 & 0.0608 & 0.0248 & 0.0320 & 0.0861 & 0.1056 & 0.0733 & 0.0795 & 0.0239 & 0.0384 & 0.0155 & 0.0201 \\
\cmidrule[0.2pt](lr){2-14}
& +BCA      & \underline{0.0409} & \underline{0.0630} & \underline{0.0266} & \underline{0.0337} & \underline{0.0887} & 0.1081 & \underline{0.0759} & \underline{0.0821} & \underline{0.0257} & 0.0405 & \textbf{0.0172} & \underline{0.0219} \\
& +IST      & 0.0393 & 0.0607 & 0.0260 & 0.0329 & 0.0871 & \underline{0.1094} & 0.0739 & 0.0810 & \textbf{0.0259} & \underline{0.0412} & \underline{0.0169} & \underline{0.0219} \\
& +IST+BCA   & \textbf{0.0423} & \textbf{0.0643} & \textbf{0.0281} & \textbf{0.0352} & \textbf{0.0909} & \textbf{0.1107} & \textbf{0.0773} & \textbf{0.0837} & \underline{0.0257} & \textbf{0.0416} & \textbf{0.0172} & \textbf{0.0223} \\
& \textbf{Improve.}  & \textbf{+10.4\%} & \textbf{+5.8\%} & \textbf{+13.3\%} & \textbf{+10.0\%} & \textbf{+5.6\%} & \textbf{+3.9\%} & \textbf{+5.5\%} & \textbf{+5.3\%} & \textbf{+4.9\%} & \textbf{+6.7\%} & \textbf{+8.2\%} & \textbf{+8.3\%} \\
\midrule
\multirow{7}{*}{TIGER-VAE}
& Fixed    & 0.0357 & 0.0532 & 0.0230 & 0.0286 & 0.0836 & 0.1018 & 0.0711 & 0.0769 & 0.0198 & 0.0330 & 0.0127 & 0.0169 \\
& SARQ    & 0.0330 & 0.0537 & 0.0213 & 0.0280 & 0.0837 & 0.1034 & 0.0710 & 0.0773 & 0.0221 & 0.0352 & 0.0145 & 0.0187 \\
& VSID  & 0.0353 & 0.0553 & 0.0227 & 0.0291 & 0.0843 & 0.1033 & 0.0718 & 0.0780 & 0.0217 & 0.0354 & 0.0144 & 0.0188 \\
\cmidrule[0.2pt](lr){2-14}
& +BCA      & \underline{0.0379} & \underline{0.0577} & \underline{0.0243} & \underline{0.0307} & 0.0863 & 0.1041 & 0.0744 & 0.0801 & \underline{0.0245} & 0.0368 & \textbf{0.0170} & \textbf{0.0209} \\
& +IST      & 0.0353 & 0.0564 & 0.0227 & 0.0295 & \underline{0.0875} & \underline{0.1067} & \underline{0.0748} & \underline{0.0810} & 0.0234 & \textbf{0.0384} & 0.0150 & 0.0198 \\
& +IST+BCA   & \textbf{0.0384} & \textbf{0.0589} & \textbf{0.0253} & \textbf{0.0319} & \textbf{0.0903} & \textbf{0.1091} & \textbf{0.0777} & \textbf{0.0837} & \textbf{0.0247} & \underline{0.0378} & \underline{0.0165} & \underline{0.0207} \\
& \textbf{Improve.}  & \textbf{+7.6\%} & \textbf{+6.5\%} & \textbf{+10.0\%} & \textbf{+9.6\%} & \textbf{+7.1\%} & \textbf{+5.5\%} & \textbf{+8.2\%} & \textbf{+7.3\%} & \textbf{+11.8\%} & \textbf{+6.8\%} & \textbf{+13.8\%} & \textbf{+10.1\%} \\
\midrule
\multirow{7}{*}{LETTER}
& Fixed    & 0.0400 & 0.0635 & 0.0264 & 0.0340 & 0.0865 & 0.1070 & 0.0738 & 0.0804 & 0.0266 & 0.0413 & 0.0175 & 0.0222 \\
& SARQ    & 0.0409 & 0.0644 & 0.0266 & 0.0341 & 0.0845 & 0.1060 & 0.0641 & 0.0711 & 0.0260 & 0.0432 & 0.0171 & 0.0226 \\
& VSID  & 0.0410 & 0.0660 & 0.0268 & 0.0348 & 0.0863 & 0.1071 & 0.0739 & 0.0806 & 0.0257 & 0.0411 & 0.0171 & 0.0220 \\
\cmidrule[0.2pt](lr){2-14}
& +BCA      & 0.0428 & 0.0651 & \underline{0.0286} & 0.0357 & \underline{0.0873} & \underline{0.1081} & \underline{0.0752} & \underline{0.0819} & 0.0263 & 0.0430 & 0.0173 & 0.0227 \\
& +IST      & \textbf{0.0444} & \textbf{0.0673} & \textbf{0.0292} & \underline{0.0366} & 0.0868 & 0.1075 & 0.0748 & 0.0815 & \underline{0.0279} & \underline{0.0438} & \underline{0.0182} & \underline{0.0233} \\
& +IST+BCA   & \underline{0.0431} & \underline{0.0669} & \textbf{0.0292} & \textbf{0.0369} & \textbf{0.0888} & \textbf{0.1085} & \textbf{0.0767} & \textbf{0.0830} & \textbf{0.0287} & \textbf{0.0452} & \textbf{0.0186} & \textbf{0.0239} \\
& \textbf{Improve.}  & \textbf{+5.1\%} & \textbf{+1.4\%} & \textbf{+9.0\%} & \textbf{+6.0\%} & \textbf{+2.7\%} & \textbf{+1.3\%} & \textbf{+3.8\%} & \textbf{+3.0\%} & \textbf{+7.9\%} & \textbf{+4.6\%} & \textbf{+6.3\%} & \textbf{+5.8\%} \\
\bottomrule
\end{tabular}
\end{table*}

We evaluate SST through extensive benchmarks on three public datasets. Following the setup and main results, we conduct a series of mechanistic analyses that dissect SST from multiple angles: attention workload verification, hyperparameter sensitivity, efficiency, coverage validation, and asymmetric design justification.

\subsection{Experimental Setup}
\label{sec:exp-setup}

\subsubsection{Datasets.}
We conduct extensive experiments on three diverse benchmark datasets: 
\textbf{Beauty} and 
\textbf{Instruments} from the Amazon product collection\footnote{\url{https://nijianmo.github.io/amazon/index.html}}, 
and \textbf{Yelp} from the business review platform\footnote{\url{https://www.yelp.com/dataset}}. Following common practice~\cite{kang2018self}, we apply 5-core filtering to remove users and items with fewer than five interactions. The remaining interactions are grouped by user and organized into chronological sequences, with the maximum sequence length truncated to 20. Table~\ref{tab:dataset-stats} summarizes the detailed statistics of the datasets after preprocessing.

\begin{table}[t]
\centering
\caption{Dataset statistics.}
\label{tab:dataset-stats}
\begin{tabular}{lrrr}
\toprule
Dataset & \#Users & \#Items & \#Interactions \\
\midrule
Beauty & 22,363 & 12,101 & 198,502 \\
Instruments & 24,772 & 9,922 & 206,153 \\
Yelp & 30,431 & 20,033 & 316,354 \\
\bottomrule
\end{tabular}
\end{table}

\subsubsection{Baseline models.}
We evaluate SST across three representative generative recommender backbones:

\begin{itemize}
\item \textbf{TIGER RQVAE}~\cite{rajput2023recommender} (denoted \textbf{TIGER-VAE}): A pioneering generative recommendation framework that employs RQ-VAE to encode item textual content into hierarchical, fixed-length semantic IDs for autoregressive retrieval.
\item \textbf{TIGER RQ-KMeans} (denoted \textbf{TIGER-KM}): An industrially aligned variant of TIGER that replaces the RQ-VAE module with Residual K-Means (RQ-KMeans)~\cite{zhou2025onerec}. As a widely adopted mainstream practice in industry, it quantizes item representations into semantic IDs via sequential clustering, serving as a highly robust encoding baseline.
\item \textbf{LETTER}~\cite{wang2024learnable}: A widely adopted improvement over TIGER that incorporates an explicit item-level alignment loss to better synchronize the generated semantic IDs with underlying collaborative signals.
\end{itemize}

Each backbone is evaluated under the following settings:

\begin{itemize}
\item \textbf{Fixed}: The original backbone with standard fixed-length SIDs, serving as the baseline.
\item \textbf{SARQ}~\cite{wang2026towards}: Truncates each item's SID at the earliest depth where the prefix bucket becomes sufficiently discriminative, using normalized entropy and dominance thresholds to determine the cut point adaptively.
\item \textbf{VSID}~\cite{khrylchenko2026variable}: Selects the optimal prefix length per item by minimizing a cost function that jointly balances entropy, collision penalty, and length regularization.
\end{itemize}

For a controlled comparison, we adapt SARQ and VSID to only modify the history-side representations while retaining fixed-length SIDs on the target side. This isolates the effect of history-side variable-length tokenization from changes in the decoding grammar, yielding controlled history-side baselines under the same target-side decoding setting as SST.  Detailed descriptions and comparisons are provided in Section~\ref{sec:related-varlen-sid}.

\begin{itemize}
\item \textbf{+BCA}: Fixed-length SIDs augmented with behavior-induced co-occurrence augmentation;
\item \textbf{+IST}: Item-level subword tokenization;
\item \textbf{+IST+BCA (SST)}: The full Semantic Subword Tokenization framework combining IST and BCA.
\end{itemize}

\subsubsection{Evaluation settings.}
To evaluate the performance of sequential recommendation, we adopt two standard metrics: top-$K$ Recall and Normalized Discounted Cumulative Gain (NDCG@$K$), where $K \in \{5, 10\}$. In alignment with existing literature~\cite{kang2018self,rajput2023recommender}, we partition the datasets using the widely adopted leave-one-out strategy. Specifically, for each user sequence, the last interacted item is held out for testing, the second-to-last for validation, and the remaining items serve as training data. To ensure a rigorous and unbiased comparison, we conduct a full-ranking evaluation across the entire item set rather than relying on sampled negatives. Furthermore, the beam size during autoregressive decoding is consistently set to 20 for all generative models.

\subsubsection{Implementation details.}

We describe the key implementation choices for the item tokenizer and the generative recommender in detail below.

\noindent\textbf{Item tokenizer.}
Following TIGER~\cite{rajput2023recommender}, we utilize Sentence-T5~\cite{ni2022sentence} to extract semantic embeddings from the textual features of each item. All three backbones adopt the Residual Quantization (RQ) framework with 4 codebook layers of size 256 codewords. For IST, we select the tokenization criterion from \{BPE, WordPiece, CondEntropy\} for each backbone--dataset combination solely according to validation loss. Table~\ref{tab:main-results} reports the test performance of the resulting fixed configuration, while Table~\ref{tab:criterion-comparison} reports the test results of all criterion-specific configurations. The subword count $k$ is tuned from $\{64, 128, 192, 256, 320, 384, 512\}$, minimum subword frequency $f$ from $\{20, 30\}$, and item support $s$ from $\{10, 15\}$. For BCA, the semantic prefix length is fixed at $P=2$ and the replay history length at $H=5$. The sliding window size $W$ is tuned from $\{3, 5\}$, minimum co-occurrence count $c$ from $\{3, 5\}$, top-$k$ prefix pairs $k$ from $\{64, 128, 256, 384, 512\}$, and the maximum number of augmented samples per sequence from $\{10, 30, 50, 100\}$.

\noindent\textbf{Generative recommender.}
All backbones share a T5-style~\cite{raffel2020exploring} encoder--decoder Transformer architecture with 4 encoder layers, 4 decoder layers, hidden size 128, FFN intermediate size 1024, and 6 attention heads. The model is optimized with AdamW at learning rate $5\times10^{-4}$, per-device batch size 256 with gradient accumulation steps 2. Training runs for up to 200 epochs with early stopping patience of 20, selecting the best checkpoint by validation loss. Both checkpoint selection and all SST configuration-selection decisions use validation loss only; the test set is evaluated only after the configuration and checkpoint are fixed. During inference, beam search with beam size 20 and full-trie constrained decoding is applied.

\subsection{Overall Performance}
\label{sec:overall-performance}

Table~\ref{tab:main-results} reports the main results. We highlight three observations.

\noindent\textbf{Backbone comparison.}
LETTER consistently achieves the strongest absolute performance across all datasets, benefiting from its explicit item-level alignment loss. Between the two TIGER variants, TIGER-KM generally outperforms TIGER-VAE due to more stable semantic partitions. Despite these capability differences, SST yields consistent relative gains across all three backbones, demonstrating its generality.

\noindent\textbf{Comparison with baselines.}
As transferable variable-length SID baselines, SARQ and VSID provide a useful test of whether shortening history-side token sequences alone is sufficient. They sometimes improve over Fixed, but the gains are dataset- and backbone-dependent, and SARQ can even underperform Fixed (e.g., on Beauty). In comparison, IST delivers more stable gains across backbone--dataset combinations, suggesting that SST benefits from semantic-subword-aware history compression rather than merely shorter sequences. Additional comparisons with length-matched random segmentation and InputPool are provided in Appendix~\ref{sec:history-side-baselines}.

\noindent\textbf{Ablation of method components.}
IST and BCA each usually improve over Fixed, and SST (+IST+BCA) achieves the best or near-best performance in most cases. Although the two components are not strictly additive in every metric, the overall trend supports our core claim: cognitive offloading via tokenization and attention focusing via behavioral guidance are complementary---the former frees attention capacity from low-level reassembly, and the latter directs it toward high-order transitions.

\subsection{Attention Overload Analysis}
\label{sec:attention-overload}

While the recall gains in Table~\ref{tab:main-results} confirm that SST improves recommendation quality, they do not directly verify our core claim that SST alleviates intra-item attention overload through IST. To further examine this mechanism, we conduct a token-level attention workload analysis. We instrument the encoder self-attention layers and measure, for every query token, how its attention mass is distributed across same-item versus cross-item key tokens. Two metrics are of central interest:

\begin{itemize}
\item \textbf{Intra-Item Attention Budget}: the total attention mass allocated to other tokens from the same item across the entire input sequence. This metric directly quantifies how much of the Transformer's limited attention budget is consumed by within-item token reassembly.
\item \textbf{Atom Reassembly Load}: the average attention mass that atom tokens assign to other atom tokens within the same item. It directly quantifies how heavily the model relies on low-level token-to-token reassembly to represent an item---the very burden SST is designed to offload through IST.
\end{itemize}

\begin{figure}[t]
\centering
\includegraphics[width=\columnwidth]{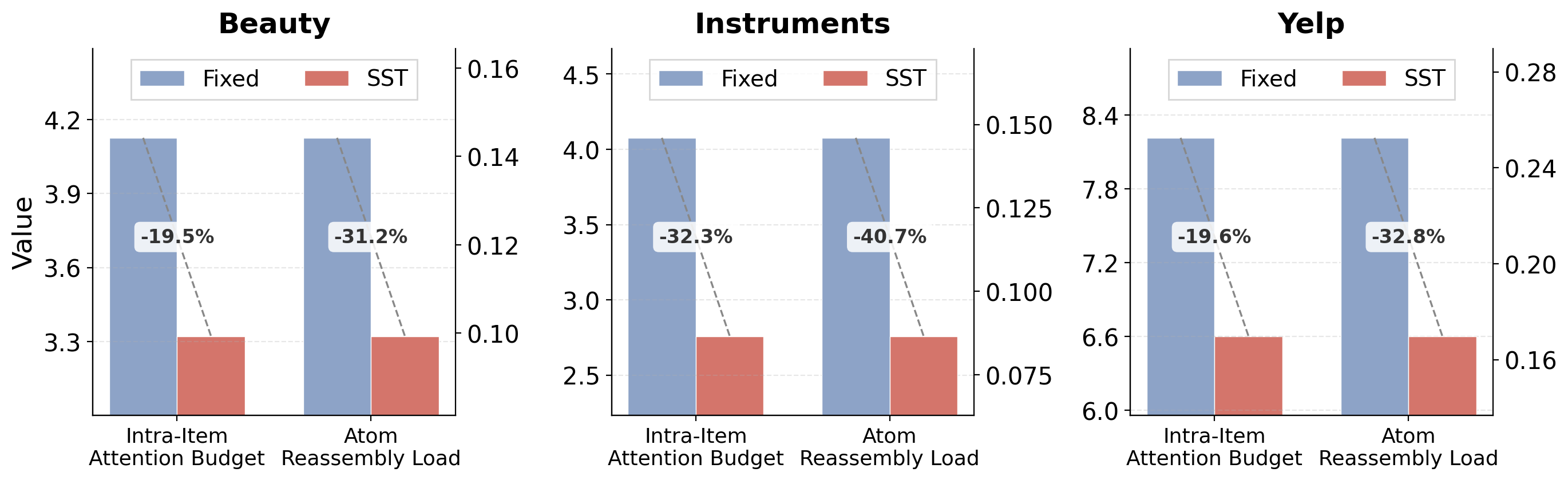}
\caption{Token-level attention workload analysis on TIGER-VAE. The two x-axis groups represent the Intra-Item Attention Budget (left axis) and Atom Reassembly Load (right axis). Dashed connectors between bar pairs indicate the relative change $\Delta$ (\%) from Fixed to SST.}
\Description{A grouped bar chart for Beauty, Instruments, and Yelp comparing Fixed SID with SST on intra-item attention budget and atom reassembly load. For every dataset and both measures, the SST bar is lower than the Fixed bar. Dashed connectors annotate the relative reduction from Fixed to SST.}
\label{fig:attention-workload}
\end{figure}

Figure~\ref{fig:attention-workload} reports the comparison between Fixed and SST across all three datasets under the TIGER-VAE backbone.

Under Fixed SID, both metrics register substantial values, reflecting the considerable attention budget consumed by intra-item token reassembly---the intra-item attention overload we hypothesized. After applying SST, both metrics drop consistently. The decline in Intra-Item Attention Budget means that the total attention mass locked up by intra-item token interactions is reduced: the model no longer needs to spend as much of its constrained budget on reassembling tokens that belong to the same item, freeing capacity for learning inter-item transition regularities---precisely the signals that BCA explicitly reinforces. The decline in Atom Reassembly Load further shows that atom queries place less attention on other atom tokens within the same item, suggesting that IST reduces the need to model atom-level co-occurrence regularities explicitly. To verify that this offloaded attention is not simply discarded, we decompose the same-item attention into sub-channels: atom$\rightarrow$subword (atom queries attending to semantic subword tokens within the same item) and subword$\rightarrow$atom (semantic subword queries attending to atom tokens). Both channels remain non-zero when SST is active, confirming that within-item representation is absorbed by semantic subword tokens as compact semantic carriers rather than eliminated.

In summary, the token-level analysis shows consistently that SST, through its IST component, reduces the attention budget allocated to low-level intra-item token reassembly, while BCA redirects the freed capacity toward inter-item co-occurrence modeling more effectively overall.

\subsection{Hyperparameter Sensitivity}
\label{sec:hyperparameter-sensitivity}

We investigate the sensitivity of the two core components to their primary hyperparameter. For IST, $k$ is the number of semantic subword merges retained in the vocabulary; for BCA, $k$ is the number of behavior prefix pairs selected for augmentation. All experiments in this section use TIGER-KM as the backbone with the best-performing tokenization criterion per dataset.

\begin{figure}[t]
\centering
\includegraphics[width=\columnwidth]{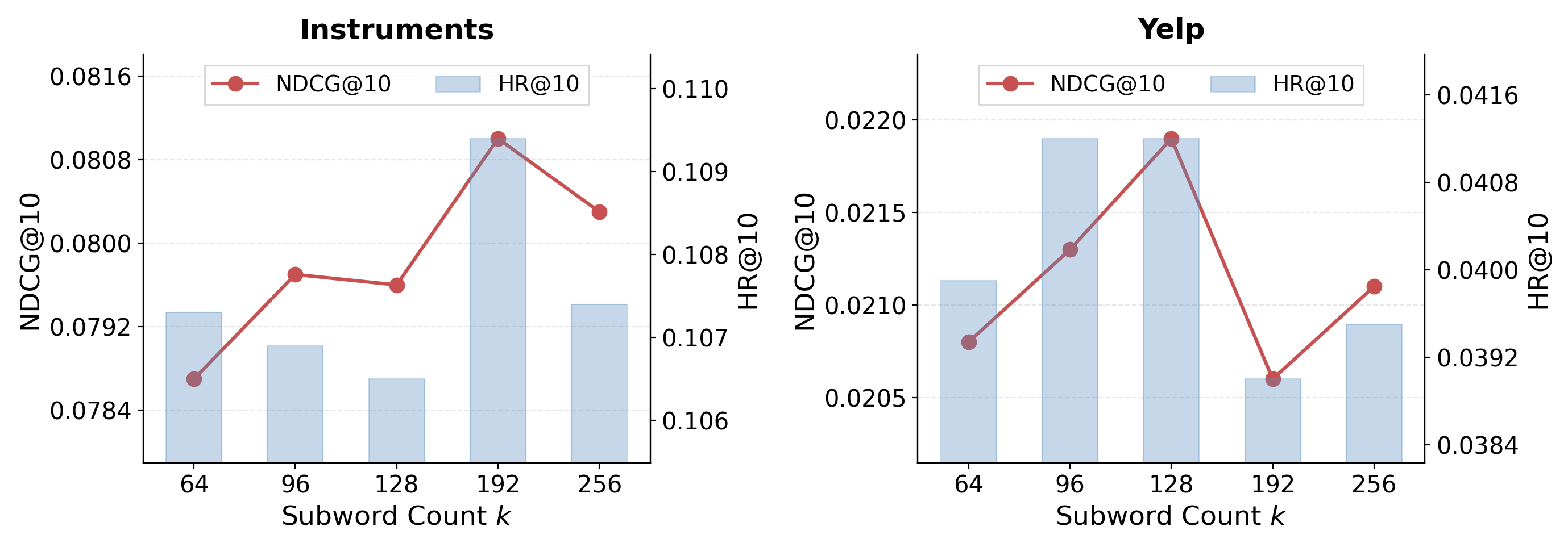}
\caption{Sensitivity of IST to the subword count $k$ on TIGER-KM. NDCG@10 shown as lines (left axis); HR@10 shown as bars (right axis).}
\Description{A dual-axis sensitivity chart for the number of IST semantic subwords on TIGER-KM. For each evaluated dataset, NDCG at 10 is plotted as a line and HR at 10 as bars across increasing subword counts. The curves peak at a moderate count and decline or flatten at larger counts.}
\label{fig:hyperparam-ist}
\end{figure}

\noindent\textbf{IST Subword Count.}
Figure~\ref{fig:hyperparam-ist} reports the effect of varying the subword count $k$. On both datasets, performance follows a concave trend: moderate $k$ works best, while overly large $k$ introduces noisy or low-confidence subwords with limited utility.

\begin{figure}[t]
\centering
\includegraphics[width=\columnwidth]{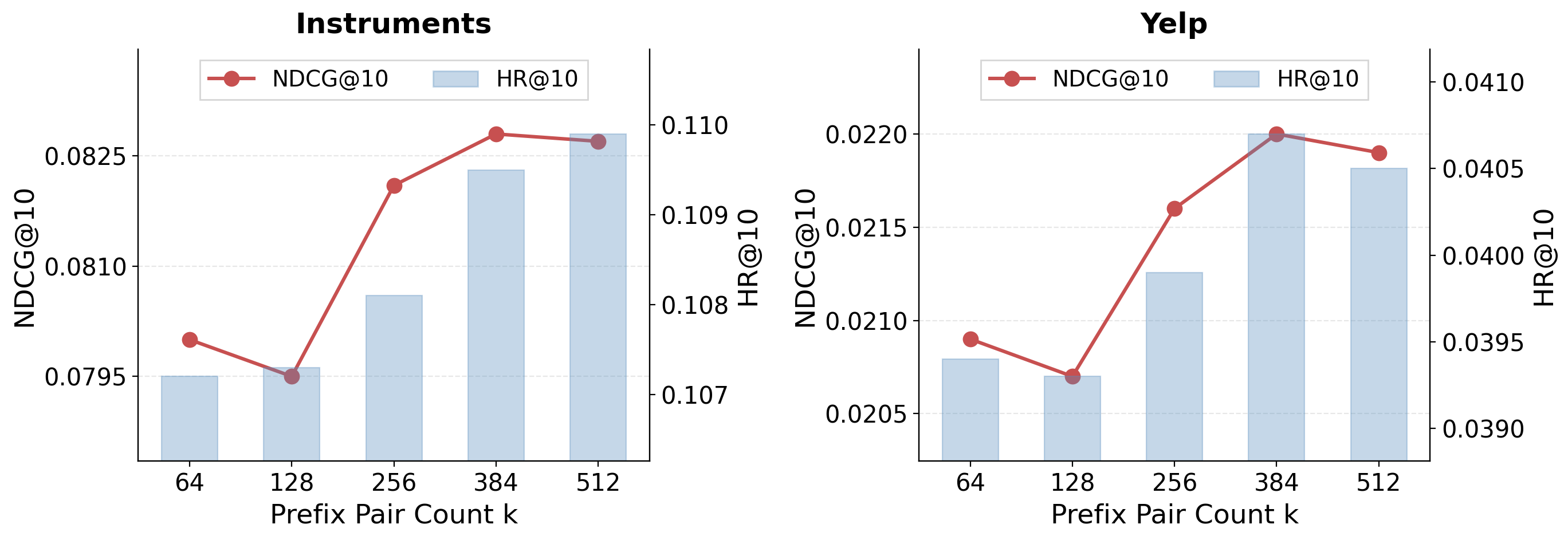}
\caption{Sensitivity of BCA to the prefix pair count $k$ on TIGER-KM. NDCG@10 shown as lines (left axis); HR@10 shown as bars (right axis).}
\Description{A dual-axis sensitivity chart for the number of BCA prefix pairs on TIGER-KM. For each evaluated dataset, NDCG at 10 is plotted as a line and HR at 10 as bars while the number of selected prefix pairs increases. Both measures improve at first and then level off at moderate pair counts.}
\label{fig:hyperparam-bca}
\end{figure}

\noindent\textbf{BCA Prefix Pair Count.}
Figure~\ref{fig:hyperparam-bca} shows the sensitivity to the prefix pair count. On both datasets, performance improves as $k$ increases and saturates at moderate values, confirming that high-quality co-occurrence transitions bring meaningful gains while further increasing $k$ introduces little additional benefit.

\subsection{Efficiency Analysis}
\label{sec:efficiency-analysis}

IST reduces the average history token length by packing multi-token spans into single semantic subword tokens, and BCA injects additional behavioral replay samples. Table~\ref{tab:efficiency} quantifies these effects alongside the end-to-end wall time (covering training, best-model loading, and test evaluation).

\begin{table}[t]
\small
\centering
\caption{End-to-end efficiency comparison on TIGER-KM. Time covers the full pipeline. History length is the average token count per test sequence. ``+Samples'' denotes the number of behavior replay samples added to the training set.}
\label{tab:efficiency}
\begin{tabular}{llrrrr}
\toprule
Dataset & Setting & Time & Epoch & Avg Hist Len & +Samples \\
\midrule
\multirow{2}{*}{Beauty}
& Fixed & 0.718 h & 60 & 28.61 & --- \\
& SST & 0.659 h & 49 & 26.73 & 11,150 \\
\midrule
\multirow{2}{*}{Instruments}
& Fixed & 0.729 h & 56 & 27.92 & --- \\
& SST & 0.691 h & 55 & 23.41 & 7,308 \\
\midrule
\multirow{2}{*}{Yelp}
& Fixed & 1.689 h & 98 & 32.53 & --- \\
& SST & 1.709 h & 92 & 29.34 & 5,753 \\
\bottomrule
\end{tabular}
\end{table}

Two opposing factors affect the overall wall time. IST reduces the average history token length by packing multi-token spans into single semantic subword tokens, which directly lowers the quadratic cost of self-attention. BCA injects additional behavior replay samples into the training set, which increases the per-epoch training cost. In all cases the two factors roughly cancel out, and the number of training epochs to convergence is consistently reduced. Overall, SST incurs no systematic slowdown despite the added IST and BCA components.

\subsection{Coverage Bucket Analysis}
\label{sec:coverage-bucket}

\begin{figure}[t]
\centering
\includegraphics[width=\columnwidth]{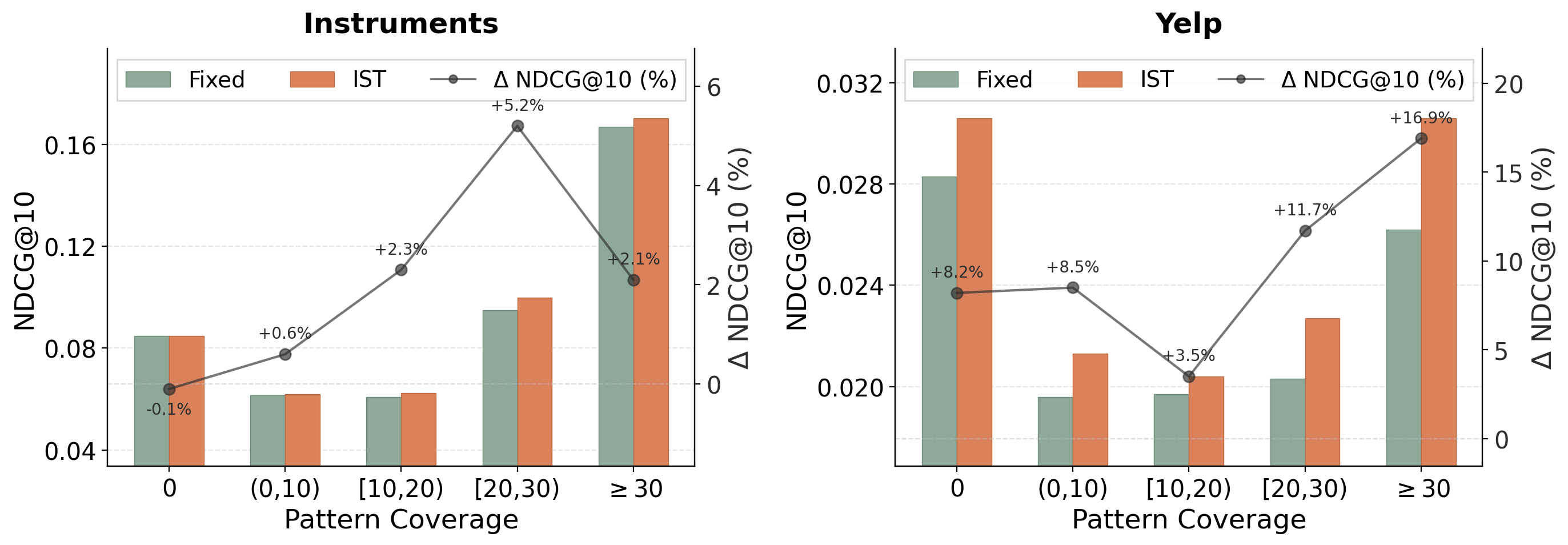}
\caption{Coverage bucket analysis on TIGER-KM. Bars show NDCG@10 for Fixed and IST per bucket; the diamond line shows the relative improvement $\Delta$NDCG@10 (\%) on the right axis.}
\Description{A coverage-bucket chart for Instruments and Yelp. Each bucket has bars for Fixed and IST NDCG at 10, while a diamond line gives the relative improvement of IST. The relative benefit generally rises as the fraction of semantic subword tokens in the history increases, with dataset-specific variation in the zero-coverage bucket.}
\label{fig:coverage-bucket}
\end{figure}

To verify that the performance gains of IST are genuinely driven by subword-aware history tokenization rather than incidental factors, we partition test samples by the fraction of semantic subword tokens ($\langle p\_*\rangle$) in their history and compare Fixed and IST within each bucket. Figure~\ref{fig:coverage-bucket} reports the results for TIGER-KM under a fine-grained bucketing scheme. The diamond line shows the relative improvement ($\Delta\%$) of IST over Fixed.

On Instruments, the relative gain is near zero in the 0\% coverage bucket and grows steadily with semantic subword coverage, peaking at moderate-to-high coverage before a natural saturation. This trend aligns with the mechanistic expectation of history-side tokenization: the more semantic subword tokens present in the history, the greater the benefit of structured subword representation. On Yelp, all coverage buckets exhibit positive relative gains, with the strongest improvement concentrated in the highest coverage tier. The 0\% coverage bucket is also positive on Yelp, whereas it is near zero on Instruments. We therefore do not interpret zero-coverage performance as direct evidence of a subword-token benefit; the more consistent pattern is that the gain increases with subword coverage across different settings.

\subsection{Asymmetric Tokenization Design}
\label{sec:asymmetric-tokenization}

A key design choice in SST is that IST and BCA only transform the encoder (history) side, while the decoder target consistently remains a fixed-length SID throughout. To validate this asymmetric design empirically, we conduct a controlled experiment on Yelp with TIGER-KM, comparing a variable-length history with a fixed-length target (our default) against variable-length histories and targets. Table~\ref{tab:fixed-vs-varlen-target} reports the results.

\begin{table}[t]
\small
\centering
\caption{Fixed target vs. VarLen target on Yelp (TIGER-KM, BPE).}
\label{tab:fixed-vs-varlen-target}
\setlength{\tabcolsep}{3.5pt}
\begin{tabular}{llrrrr}
\toprule
History & Target & HR@5 & HR@10 & NDCG@5 & NDCG@10 \\
\midrule
Fixed & Fixed & 0.0244 & 0.0385 & 0.0159 & 0.0205 \\
VarLen & Fixed & \textbf{0.0259} & \textbf{0.0412} & \textbf{0.0169} & \textbf{0.0219} \\
VarLen & VarLen & 0.0216 & 0.0329 & 0.0142 & 0.0178 \\
\bottomrule
\end{tabular}
\end{table}

Switching to a variable target causes a substantial drop, even though the same variable-length representation improves performance when used only on the history side. This contrast suggests that the difficulty is not the semantic subwords themselves, but their role as autoregressive target symbols. 

To diagnose the cause, we partition test items by whether their SID would contain at least one semantic subword token. Table~\ref{tab:merged-unmerged-bucket} shows the per-bucket results.

\begin{table}[t]
\small
\centering
\caption{Merged vs.\ Unmerged target bucket analysis on Yelp (TIGER-KM, BPE).}
\label{tab:merged-unmerged-bucket}
\begin{tabular}{llrrr}
\toprule
History & Target & Bucket & HR@10 & NDCG@10 \\
\midrule
Fixed & Fixed & merged & 0.0405 & 0.0210 \\
Fixed & Fixed & unmerged & 0.0381 & 0.0202 \\
\midrule
VarLen & Fixed & merged & 0.0446 & 0.0230 \\
VarLen & Fixed & unmerged & 0.0391 & 0.0212 \\
\midrule
VarLen & VarLen & merged & 0.0002 & 0.0001 \\
VarLen & VarLen & unmerged & 0.0530 & 0.0288 \\
\bottomrule
\end{tabular}
\end{table}

With a variable-length history and fixed-length target, both merged and unmerged items improve, confirming that merged items are not inherently harder to predict. Once the target side is also converted to a variable-length representation, however, the merged bucket collapses to near-zero performance, while the unmerged bucket remains functional. This is because unmerged items still follow the familiar 4-slot <a><b><c><d> grammar, whereas merged items require the decoder to generate semantic subword tokens or shorter SIDs during autoregressive decoding.

Further diagnostics show that semantic subword target categories uniformly degrade under a variable target. At early generation positions, semantic subword tokens rarely enter the top beams and are crowded out by dominant atom tokens. The underlying cause is training-data imbalance: semantic subword tokens form a small fraction of position-level targets, making them long-tail decoder symbols that the model reliably avoids.

Overall, variable-length histories with fixed-length targets preserve structured context and a stable 4-slot decoding grammar.

\section{Related Work}
\label{sec:related}
\sloppy

\subsection{Generative Recommendation}

Generative recommendation reformulates the next-item prediction task into an autoregressive sequence generation problem, where items are represented as discrete token sequences~\cite{wang2025generative,wu2024survey}. Existing methodologies for constructing these item identifiers can be categorized into text-based methods and SID-based methods. 

Specifically, text-based methods~\cite{geng2022recommendation,cui2022m6,tan2024idgenrec} use item descriptions, such as titles and categories, as inputs to standard language models. While leveraging LLMs' linguistic priors, they often suffer from long sequences and ambiguous item grounding during decoding.

To circumvent the limitations of raw text tokens, another dominant stream pivots toward SID-based paradigms. The most widely adopted approach is residual-based quantization, which constructs a coarse-to-fine representation by quantizing the residual between the latent embedding and the cluster centroid. Specifically, Residual Quantization VAE (RQ-VAE) serves as the foundational architecture for frameworks like TIGER~\cite{rajput2023recommender}, LC-Rec~\cite{zheng2024adapting}, and COBRA~\cite{yang2026sparse}. To prevent codebook collapse during training, subsequent variants such as OneRec~\cite{zhou2025onerec,zhou2025onerec2}, OneLoc~\cite{wei2026oneloc}, CAR~\cite{wang2025act} and GPR~\cite{zhang2025gpr} adopt Residual K-Means (RQ-KMeans) to enforce strict cluster constraints. 
Beyond basic SID construction, recent work has explored several related directions, including collision-aware codebook learning and item-level disambiguation~\cite{hu2026stop,fu2025forge,zhang2025purely}, collaborative~\cite{wang2024learnable,wang2024eager,hong2025eager,xiao2025unger} or multimodal~\cite{xu2026mmq,wang2025empowering,kimtoken} semantic alignment, and end-to-end optimization that reduces the mismatch between SID construction and generative recommendation training~\cite{liu2025generative,li2026unigrec,fu2026differentiable}.

The broader generative-recommendation literature has also studied scalable architectures~\cite{ye2026fuxilinear,ye2025fuxialpha,ye2025fuxibeta,pan2025revisiting}, empirical scaling behavior~\cite{guo2024scaling,shen2025plaw}, and unified generative retrieval and ranking~\cite{zhang2025killing}. Complementary efforts improve sequential recommendation through data regeneration or distillation~\cite{yin2024dr4sr,zhang2025td3}, diffusion-based sequence modeling~\cite{xie2024fuzzymodeling}, and efficient lifelong or multi-behavior modeling~\cite{zhou2026survey,wang2024denoising}; recent work also extends recommenders toward reasoning, self-improvement, and agentic interaction~\cite{zhang2026whythinking,zhang2026can,yu2026planagent,zhang2026useragent}.

However, existing SID paradigms rely on fixed-length tokenization, which forces the recommendation backbone to spend excessive attention on low-level intra-item token dependencies. This attention overload limits its ability to model high-level user preference transitions.

\subsection{Variable-Length Semantic IDs}
\label{sec:related-varlen-sid}

Recent studies have explored variable-length Semantic IDs as an alternative to fixed-length item codes, allowing different items to be represented with different code lengths.

\textbf{SARQ}~\cite{wang2026towards} dynamically assigns code lengths based on item-level path entropy. It traverses the RQ codebook hierarchy and truncates each SID once the current prefix becomes sufficiently discriminative, as measured by normalized entropy and dominance ratio. Its key design principle is to assign \emph{longer, more discriminative IDs to head items} that require fine-grained differentiation, and \emph{shorter, generalizable IDs to tail items} where many-way distinction is rarely required in practice.

\textbf{VSID}~\cite{khrylchenko2026variable} approaches variable-length ID construction from the perspective of emergent communication. It formulates SID learning as a discrete variational autoencoding problem with Gumbel-Softmax reparameterization, where the tokenizer learns to emit variable-length codes under reconstruction, collision, and length regularization. This framework naturally assigns \emph{compact codes to popular items} and \emph{longer codes to rare items}, mirroring the abbreviation principle observed in natural language.

These variable-length SID methods address a different problem from SST. Their focus is item-side identifier allocation: SARQ and VSID adapt code length according to item-level discriminativeness or popularity. In contrast, SST studies history-side attention overload, where flattened fixed-length SIDs force the model to repeatedly reassemble atom tokens into item-level signals. We therefore treat SARQ and VSID as transferable variable-length SID baselines: their encoding rules can be adapted to the history side to test whether merely shortening history-side token sequences is sufficient, but they are not designed to solve the same attention-overload problem we study here.

\subsection{Tokenization in Large Language Models}

Tokenization is a fundamental component in LLMs that defines the basic units of representation. Modern NLP commonly adopts subword tokenizers such as BPE~\cite{sennrich2016neural}, WordPiece~\cite{devlin2019bert}, and SentencePiece/Unigram~\cite{kudo2018sentencepiece,kudo2018subword}, which balance vocabulary size and sequence length while alleviating the out-of-vocabulary problem~\cite{mielke2021between}. Recent studies further show that tokenization affects not only compression efficiency but also downstream model behavior and learning dynamics~\cite{goldman2024unpacking,rajaraman2024toward,erdogan2026information}.

Inspired by these insights, SST adapts subword tokenization to generative recommendation. Unlike fixed-length SIDs, which resemble character-level spelling, SST merges stable co-occurrences into compact semantic subword tokens, reducing intra-item redundancy and freeing the backbone to model higher-level regularities.

\section{Conclusion}
\label{sec:conclusion}

This paper identifies a user-context tokenization bottleneck in SID-based generative recommendation: fixed-length SIDs are stable target identifiers, but when flattened to represent user histories, they consume attention on low-level intra-item atom reassembly. We propose Semantic Subword Tokenization (SST), which addresses this bottleneck through two complementary components. IST merges stable adjacent atom co-occurrences into reusable semantic subword tokens, offloading low-level intra-item regularities from the Transformer. BCA injects coarse-grained behavioral co-occurrence signals, redirecting the freed capacity toward higher-order inter-item transitions. SST further adopts asymmetric tokenization by retaining fixed-length SIDs on the target side to preserve decoding stability. Experiments across three datasets and three generative recommender backbones show that SST improves over fixed-length and transferable variable-length SID baselines.

\appendix

\section{Tokenization Criterion Comparison}
\label{sec:criterion-comparison}
\sloppy

IST is compatible with multiple tokenization criteria. Table~\ref{tab:criterion-comparison} compares BPE, WordPiece, and CondEntropy across all backbone--dataset combinations.

\begin{table*}[t]
\fontsize{8.5pt}{9.8pt}\selectfont
\centering
\caption{Comparison of tokenization criteria. N@5/N@10 abbreviate NDCG@5/NDCG@10.}
\label{tab:criterion-comparison}
\setlength{\tabcolsep}{3.4pt}
\renewcommand{\arraystretch}{1.08}
\begin{tabular}{ll|rrrr|rrrr|rrrr}
\toprule
& & \multicolumn{4}{c|}{Beauty} & \multicolumn{4}{c|}{Instruments} & \multicolumn{4}{c}{Yelp} \\
\cmidrule(lr){3-6} \cmidrule(lr){7-10} \cmidrule(lr){11-14}
Backbone & Setting & HR@5 & HR@10 & N@5 & N@10 & HR@5 & HR@10 & N@5 & N@10 & HR@5 & HR@10 & N@5 & N@10 \\
\midrule
\multirow{4}{*}{TIGER-KM}
& Fixed & 0.0372 & 0.0582 & 0.0238 & 0.0305 & 0.0849 & 0.1065 & 0.0718 & 0.0787 & 0.0244 & 0.0385 & 0.0159 & 0.0205 \\
& BPE & \textbf{0.0393} & 0.0607 & \textbf{0.0260} & \textbf{0.0329} & \textbf{0.0874} & 0.1070 & 0.0738 & 0.0801 & \textbf{0.0259} & \textbf{0.0412} & \textbf{0.0169} & \textbf{0.0219} \\
& WordPiece & 0.0389 & 0.0609 & 0.0252 & 0.0323 & 0.0867 & 0.1089 & 0.0736 & 0.0808 & 0.0248 & 0.0395 & 0.0161 & 0.0209 \\
& CondEntropy & 0.0386 & \textbf{0.0621} & 0.0247 & 0.0322 & 0.0871 & \textbf{0.1094} & \textbf{0.0739} & \textbf{0.0810} & 0.0247 & 0.0394 & 0.0162 & 0.0209 \\
\midrule
\multirow{4}{*}{TIGER-VAE}
& Fixed & \textbf{0.0357} & 0.0532 & \textbf{0.0230} & 0.0286 & 0.0836 & 0.1018 & 0.0711 & 0.0769 & 0.0198 & 0.0330 & 0.0127 & 0.0169 \\
& BPE & 0.0353 & 0.0564 & 0.0227 & \textbf{0.0295} & \textbf{0.0875} & \textbf{0.1067} & \textbf{0.0748} & \textbf{0.0810} & 0.0226 & 0.0355 & 0.0144 & 0.0185 \\
& WordPiece & 0.0355 & 0.0553 & \textbf{0.0230} & 0.0294 & 0.0861 & 0.1054 & 0.0731 & 0.0793 & 0.0226 & 0.0366 & 0.0145 & 0.0190 \\
& CondEntropy & 0.0354 & \textbf{0.0565} & 0.0224 & 0.0292 & 0.0852 & 0.1049 & 0.0727 & 0.0790 & \textbf{0.0234} & \textbf{0.0384} & \textbf{0.0150} & \textbf{0.0198} \\
\midrule
\multirow{4}{*}{LETTER}
& Fixed & 0.0400 & 0.0635 & 0.0264 & 0.0340 & 0.0865 & 0.1070 & 0.0738 & 0.0804 & 0.0266 & 0.0413 & 0.0175 & 0.0222 \\
& BPE & 0.0423 & 0.0642 & 0.0288 & 0.0358 & 0.0868 & 0.1083 & 0.0744 & 0.0813 & 0.0272 & 0.0432 & 0.0177 & 0.0228 \\
& WordPiece & \textbf{0.0444} & \textbf{0.0673} & \textbf{0.0292} & \textbf{0.0366} & \textbf{0.0872} & \textbf{0.1085} & 0.0742 & 0.0811 & 0.0272 & 0.0429 & 0.0177 & 0.0227 \\
& CondEntropy & 0.0405 & 0.0657 & 0.0270 & 0.0351 & 0.0868 & 0.1075 & \textbf{0.0748} & \textbf{0.0815} & \textbf{0.0279} & \textbf{0.0438} & \textbf{0.0182} & \textbf{0.0233} \\
\bottomrule
\end{tabular}
\end{table*}

On Beauty, WordPiece achieves the best results across most backbone--metric combinations, sweeping all metrics for LETTER. On Yelp, CondEntropy dominates across the majority of settings, and the relative gains over Fixed are substantially larger than on the other two datasets. On Instruments, no single criterion dominates: the best choice varies across backbones and metrics, and the margins over Fixed are generally smaller. This pattern suggests that both the benefit of SST and the preferred criterion depend on dataset-specific item co-occurrence structure. PMI-based scoring is effective for regular relations, whereas stricter bidirectional conditional-entropy filtering is more robust for diverse or noisy interactions. When co-occurrence structure is less pronounced, all criteria still yield net-positive gains in most settings, but both the overall margin and criterion-wise differences shrink. BPE remains competitive across settings but is not uniformly preferred, confirming that the benefit comes from history-side subword tokenization rather than a single scoring rule. These results motivate selecting the criterion separately for each backbone--dataset combination.

\section{History-Side Representation Baselines}
\label{sec:history-side-baselines}

\begin{table}[t]
\small
\centering
\caption{History-side representation baseline comparison on TIGER-KM.}
\label{tab:history-baselines}
\setlength{\tabcolsep}{1.7pt}
\begin{tabular}{l|rr|rr|rr}
\toprule
& \multicolumn{2}{c|}{Beauty} & \multicolumn{2}{c|}{Instruments} & \multicolumn{2}{c}{Yelp} \\
Setting & HR@10 & NDCG@10 & HR@10 & NDCG@10 & HR@10 & NDCG@10 \\
\midrule
Fixed & 0.0582 & 0.0305 & 0.1065 & 0.0787 & 0.0385 & 0.0205 \\
IST & \textbf{0.0607} & \textbf{0.0329} & \textbf{0.1094} & \textbf{0.0810} & \textbf{0.0412} & \textbf{0.0219} \\
InputPool & 0.0185 & 0.0092 & 0.0473 & 0.0230 & 0.0105 & 0.0055 \\
Random & 0.0593 & 0.0311 & 0.1051 & 0.0787 & 0.0352 & 0.0182 \\
\bottomrule
\end{tabular}
\end{table}

Table~\ref{tab:history-baselines} compares IST with InputPool and length-matched Random on TIGER-KM, both retaining the fixed target SID and decoding trie. The two alternatives isolate different simplifications of the history representation. InputPool replaces the ordered atom sequence of each item with its mean embedding, removing atom-level sequential composition altogether. Random uses contiguous atom spans with lengths matched to the IST reference tokenization, preserving comparable compression while discarding data-derived merge structure. IST consistently outperforms Fixed and both alternatives. Its advantage over Random indicates that shortening histories alone is insufficient, while its advantage over InputPool shows that compact learned token units retain useful structure that is lost under simple item-level pooling. Since all settings share the same target SID and decoding trie, these differences arise solely from their history-side representations.

\section{Popularity Analysis}

\begin{table}[t]
\small
\centering
\caption{Popularity analysis on Yelp (TIGER-KM).}
\label{tab:bca-popularity}
\setlength{\tabcolsep}{2.4pt}
\begin{tabular}{l|rr|rrr}
\toprule
Setting & \shortstack{Head\\top-1} & \shortstack{Tail\\top-1} & \shortstack{Head\\N@10} & \shortstack{Mid\\N@10} & \shortstack{Tail\\N@10} \\
\midrule
Fixed & 95.08\% & 0.85\% & 0.0364 & 0.0065 & 0.0045 \\
+IST & 95.56\% & 1.00\% & \textbf{0.0388} & 0.0065 & 0.0051 \\
+IST+BCA & 92.24\% & \textbf{1.67\%} & 0.0365 & \textbf{0.0104} & \textbf{0.0074} \\
\bottomrule
\end{tabular}
\end{table}

To assess whether BCA exacerbates popularity bias, we define head, mid, and tail items as the top 20\%, next 30\%, and bottom 50\% ranked by unaugmented training frequency. Table~\ref{tab:bca-popularity} reports top-1 recommendation exposure and target-bucket NDCG@10 for Yelp with TIGER-KM. Relative to Fixed, IST maintains comparable head exposure while increasing tail exposure and improving head- and tail-target NDCG@10. Thus, IST does not make recommendation exposure more concentrated on head items in this evaluated setting. Adding BCA further lowers head exposure, increases tail exposure, and improves mid- and tail-target NDCG@10, indicating that BCA partially improves long-tail exposure and recommendation quality.

\begin{acks}
This work was supported by the National Natural Science Foundation of China (Nos. 62472394, U23A20319, 62441239, and 62441227), as well as the Anhui Province Science and Technology Innovation Project (Nos. 202423k09020010 and 202423k09020011).
\end{acks}

\section*{GenAI Usage Disclosure}
During the preparation of this work, we used GPT-5.5 to improve the readability and language of the manuscript, as well as to assist in drafting specific code snippets for experimental implementation. After using this tool, we tested and verified all generated code, reviewed and edited the manuscript content, and took full responsibility for the accuracy and originality of the final publication.

\nocite{wang2025mfgslae,wang2019mcne,wang2021hypersorec,zhang2025unified,xu2025mirrn,wang2026mec,wang2025dlf,zhou2025mit,wang2025decorrelated,gu2025rapid,yin2025genctr,huang2025chemeval,xie2025bottleneck,zhang2025ragigbench,huang2025selfaug,liang2025adaptive}

\bibliographystyle{ACM-Reference-Format}
\bibliography{refs_camera_ready,refs_group_normal,refs_group_nocite}


\begin{thebibliography}{87}


\ifx \showCODEN    \undefined \def \showCODEN     #1{\unskip}     \fi
\ifx \showISBNx    \undefined \def \showISBNx     #1{\unskip}     \fi
\ifx \showISBNxiii \undefined \def \showISBNxiii  #1{\unskip}     \fi
\ifx \showISSN     \undefined \def \showISSN      #1{\unskip}     \fi
\ifx \showLCCN     \undefined \def \showLCCN      #1{\unskip}     \fi
\ifx \shownote     \undefined \def \shownote      #1{#1}          \fi
\ifx \showarticletitle \undefined \def \showarticletitle #1{#1}   \fi
\ifx \showURL      \undefined \def \showURL       {\relax}        \fi
\providecommand\bibfield[2]{#2}
\providecommand\bibinfo[2]{#2}
\providecommand\natexlab[1]{#1}
\providecommand\showeprint[2][]{arXiv:#2}

\bibitem[Achiam et~al\mbox{.}(2023)]%
        {achiam2023gpt}
\bibfield{author}{\bibinfo{person}{Josh Achiam}, \bibinfo{person}{Steven
  Adler}, \bibinfo{person}{Sandhini Agarwal}, \bibinfo{person}{Lama Ahmad},
  \bibinfo{person}{Ilge Akkaya}, \bibinfo{person}{Florencia~Leoni Aleman},
  \bibinfo{person}{Diogo Almeida}, \bibinfo{person}{Janko Altenschmidt},
  \bibinfo{person}{Sam Altman}, \bibinfo{person}{Shyamal Anadkat},
  {et~al\mbox{.}}} \bibinfo{year}{2023}\natexlab{}.
\newblock \showarticletitle{Gpt-4 technical report}.
\newblock \bibinfo{journal}{\emph{arXiv preprint arXiv:2303.08774}}
  (\bibinfo{year}{2023}).
\newblock


\bibitem[Cui et~al\mbox{.}(2022)]%
        {cui2022m6}
\bibfield{author}{\bibinfo{person}{Zeyu Cui}, \bibinfo{person}{Jianxin Ma},
  \bibinfo{person}{Chang Zhou}, \bibinfo{person}{Jingren Zhou}, {and}
  \bibinfo{person}{Hongxia Yang}.} \bibinfo{year}{2022}\natexlab{}.
\newblock \showarticletitle{M6-rec: Generative pretrained language models are
  open-ended recommender systems}.
\newblock \bibinfo{journal}{\emph{arXiv preprint arXiv:2205.08084}}
  (\bibinfo{year}{2022}).
\newblock


\bibitem[Devlin et~al\mbox{.}(2019)]%
        {devlin2019bert}
\bibfield{author}{\bibinfo{person}{Jacob Devlin}, \bibinfo{person}{Ming-Wei
  Chang}, \bibinfo{person}{Kenton Lee}, {and} \bibinfo{person}{Kristina
  Toutanova}.} \bibinfo{year}{2019}\natexlab{}.
\newblock \showarticletitle{Bert: Pre-training of deep bidirectional
  transformers for language understanding}. In
  \bibinfo{booktitle}{\emph{Proceedings of the 2019 conference of the North
  American chapter of the association for computational linguistics: human
  language technologies, volume 1 (long and short papers)}}.
  \bibinfo{publisher}{Association for Computational Linguistics},
  \bibinfo{address}{Minneapolis, MN, USA}, \bibinfo{pages}{4171--4186}.
\newblock


\bibitem[Erdogan et~al\mbox{.}(2026)]%
        {erdogan2026information}
\bibfield{author}{\bibinfo{person}{Mete Erdogan}, \bibinfo{person}{Abhiram
  Gorle}, \bibinfo{person}{Shubham Chandak}, \bibinfo{person}{Mert Pilanci},
  {and} \bibinfo{person}{Tsachy Weissman}.} \bibinfo{year}{2026}\natexlab{}.
\newblock \showarticletitle{An Information-Theoretic Perspective on LLM
  Tokenizers}.
\newblock \bibinfo{journal}{\emph{arXiv preprint arXiv:2601.09039}}
  (\bibinfo{year}{2026}).
\newblock


\bibitem[Floridi and Chiriatti(2020)]%
        {floridi2020gpt}
\bibfield{author}{\bibinfo{person}{Luciano Floridi} {and}
  \bibinfo{person}{Massimo Chiriatti}.} \bibinfo{year}{2020}\natexlab{}.
\newblock \showarticletitle{GPT-3: Its nature, scope, limits, and
  consequences}.
\newblock \bibinfo{journal}{\emph{Minds and machines}} \bibinfo{volume}{30},
  \bibinfo{number}{4} (\bibinfo{year}{2020}), \bibinfo{pages}{681--694}.
\newblock


\bibitem[Fu et~al\mbox{.}(2026)]%
        {fu2026differentiable}
\bibfield{author}{\bibinfo{person}{Junchen Fu}, \bibinfo{person}{Xuri Ge},
  \bibinfo{person}{Alexandros Karatzoglou}, \bibinfo{person}{Ioannis Arapakis},
  \bibinfo{person}{Suzan Verberne}, \bibinfo{person}{Joemon~M Jose}, {and}
  \bibinfo{person}{Zhaochun Ren}.} \bibinfo{year}{2026}\natexlab{}.
\newblock \showarticletitle{Differentiable Semantic ID for Generative
  Recommendation}.
\newblock \bibinfo{journal}{\emph{arXiv preprint arXiv:2601.19711}}
  (\bibinfo{year}{2026}).
\newblock


\bibitem[Fu et~al\mbox{.}(2025)]%
        {fu2025forge}
\bibfield{author}{\bibinfo{person}{Kairui Fu}, \bibinfo{person}{Tao Zhang},
  \bibinfo{person}{Shuwen Xiao}, \bibinfo{person}{Ziyang Wang},
  \bibinfo{person}{Xinming Zhang}, \bibinfo{person}{Chenchi Zhang},
  \bibinfo{person}{Yuliang Yan}, \bibinfo{person}{Junjun Zheng},
  \bibinfo{person}{Yu Li}, \bibinfo{person}{Zhihong Chen}, {et~al\mbox{.}}}
  \bibinfo{year}{2025}\natexlab{}.
\newblock \showarticletitle{Forge: Forming semantic identifiers for generative
  retrieval in industrial datasets}.
\newblock \bibinfo{journal}{\emph{arXiv preprint arXiv:2509.20904}}
  (\bibinfo{year}{2025}).
\newblock


\bibitem[Geng et~al\mbox{.}(2022)]%
        {geng2022recommendation}
\bibfield{author}{\bibinfo{person}{Shijie Geng}, \bibinfo{person}{Shuchang
  Liu}, \bibinfo{person}{Zuohui Fu}, \bibinfo{person}{Yingqiang Ge}, {and}
  \bibinfo{person}{Yongfeng Zhang}.} \bibinfo{year}{2022}\natexlab{}.
\newblock \showarticletitle{Recommendation as language processing (rlp): A
  unified pretrain, personalized prompt \& predict paradigm (p5)}. In
  \bibinfo{booktitle}{\emph{Proceedings of the 16th ACM conference on
  recommender systems}}. \bibinfo{publisher}{Association for Computing
  Machinery}, \bibinfo{address}{New York, NY, USA}, \bibinfo{pages}{299--315}.
\newblock


\bibitem[Goldman et~al\mbox{.}(2024)]%
        {goldman2024unpacking}
\bibfield{author}{\bibinfo{person}{Omer Goldman}, \bibinfo{person}{Avi
  Caciularu}, \bibinfo{person}{Matan Eyal}, \bibinfo{person}{Kris Cao},
  \bibinfo{person}{Idan Szpektor}, {and} \bibinfo{person}{Reut Tsarfaty}.}
  \bibinfo{year}{2024}\natexlab{}.
\newblock \showarticletitle{Unpacking tokenization: Evaluating text compression
  and its correlation with model performance}. In
  \bibinfo{booktitle}{\emph{Findings of the Association for Computational
  Linguistics: ACL 2024}}. \bibinfo{publisher}{Association for Computational
  Linguistics}, \bibinfo{address}{Bangkok, Thailand},
  \bibinfo{pages}{2274--2286}.
\newblock


\bibitem[Gu et~al\mbox{.}(2025)]%
        {gu2025rapid}
\bibfield{author}{\bibinfo{person}{Hongchao Gu}, \bibinfo{person}{Dexun Li},
  \bibinfo{person}{Kuicai Dong}, \bibinfo{person}{Hao Zhang},
  \bibinfo{person}{Hang Lv}, \bibinfo{person}{Hao Wang}, \bibinfo{person}{Defu
  Lian}, \bibinfo{person}{Yong Liu}, {and} \bibinfo{person}{Enhong Chen}.}
  \bibinfo{year}{2025}\natexlab{}.
\newblock \showarticletitle{RAPID: Efficient Retrieval-Augmented Long Text
  Generation with Writing Planning and Information Discovery}. In
  \bibinfo{booktitle}{\emph{Findings of the Association for Computational
  Linguistics: ACL 2025}}. \bibinfo{pages}{16742--16763}.
\newblock
\href{https://doi.org/10.18653/v1/2025.findings-acl.859}{doi:\nolinkurl{10.18653/v1/2025.findings-acl.859}}


\bibitem[Guo et~al\mbox{.}(2024)]%
        {guo2024scaling}
\bibfield{author}{\bibinfo{person}{Wei Guo}, \bibinfo{person}{Hao Wang},
  \bibinfo{person}{Luankang Zhang}, \bibinfo{person}{Jin~Yao Chin},
  \bibinfo{person}{Zhongzhou Liu}, \bibinfo{person}{Kai Cheng},
  \bibinfo{person}{Qiushi Pan}, \bibinfo{person}{Yi~Quan Lee},
  \bibinfo{person}{Wanqi Xue}, \bibinfo{person}{Tingjia Shen}, {et~al\mbox{.}}}
  \bibinfo{year}{2024}\natexlab{}.
\newblock \showarticletitle{Scaling New Frontiers: Insights into Large
  Recommendation Models}.
\newblock \bibinfo{journal}{\emph{arXiv preprint arXiv:2412.00714}}
  (\bibinfo{year}{2024}).
\newblock


\bibitem[Hidasi et~al\mbox{.}(2015)]%
        {hidasi2015session}
\bibfield{author}{\bibinfo{person}{Bal{\'a}zs Hidasi},
  \bibinfo{person}{Alexandros Karatzoglou}, \bibinfo{person}{Linas Baltrunas},
  {and} \bibinfo{person}{Domonkos Tikk}.} \bibinfo{year}{2015}\natexlab{}.
\newblock \showarticletitle{Session-based recommendations with recurrent neural
  networks}.
\newblock \bibinfo{journal}{\emph{arXiv preprint arXiv:1511.06939}}
  (\bibinfo{year}{2015}).
\newblock


\bibitem[Hong et~al\mbox{.}(2025)]%
        {hong2025eager}
\bibfield{author}{\bibinfo{person}{Minjie Hong}, \bibinfo{person}{Yan Xia},
  \bibinfo{person}{Zehan Wang}, \bibinfo{person}{Jieming Zhu},
  \bibinfo{person}{Ye Wang}, \bibinfo{person}{Sihang Cai},
  \bibinfo{person}{Xiaoda Yang}, \bibinfo{person}{Quanyu Dai},
  \bibinfo{person}{Zhenhua Dong}, \bibinfo{person}{Zhimeng Zhang},
  {et~al\mbox{.}}} \bibinfo{year}{2025}\natexlab{}.
\newblock \showarticletitle{Eager-llm: Enhancing large language models as
  recommenders through exogenous behavior-semantic integration}. In
  \bibinfo{booktitle}{\emph{Proceedings of the ACM on Web Conference 2025}}.
  \bibinfo{publisher}{Association for Computing Machinery},
  \bibinfo{address}{New York, NY, USA}, \bibinfo{pages}{2754--2762}.
\newblock


\bibitem[Hou et~al\mbox{.}(2025)]%
        {hou2025survey}
\bibfield{author}{\bibinfo{person}{Min Hou}, \bibinfo{person}{Le Wu},
  \bibinfo{person}{Yuxin Liao}, \bibinfo{person}{Yonghui Yang},
  \bibinfo{person}{Zhen Zhang}, \bibinfo{person}{Changlong Zheng},
  \bibinfo{person}{Han Wu}, {and} \bibinfo{person}{Richang Hong}.}
  \bibinfo{year}{2025}\natexlab{}.
\newblock \showarticletitle{A survey on generative recommendation: Data, model,
  and tasks}.
\newblock \bibinfo{journal}{\emph{arXiv preprint arXiv:2510.27157}}
  (\bibinfo{year}{2025}).
\newblock


\bibitem[Hu et~al\mbox{.}(2026)]%
        {hu2026stop}
\bibfield{author}{\bibinfo{person}{Zheng Hu}, \bibinfo{person}{Yuxin Chen},
  \bibinfo{person}{Yongsen Pan}, \bibinfo{person}{Xu Yuan},
  \bibinfo{person}{Yuting Yin}, \bibinfo{person}{Daoyuan Wang},
  \bibinfo{person}{Boyang Xia}, \bibinfo{person}{Zefei Luo},
  \bibinfo{person}{Hongyang Wang}, \bibinfo{person}{Songhao Ni},
  {et~al\mbox{.}}} \bibinfo{year}{2026}\natexlab{}.
\newblock \showarticletitle{Stop Treating Collisions Equally:
  Qualification-Aware Semantic ID Learning for Recommendation at Industrial
  Scale}.
\newblock \bibinfo{journal}{\emph{arXiv preprint arXiv:2603.00632}}
  (\bibinfo{year}{2026}).
\newblock


\bibitem[Huang et~al\mbox{.}(2025a)]%
        {huang2025chemeval}
\bibfield{author}{\bibinfo{person}{Yuqing Huang}, \bibinfo{person}{Rongyang
  Zhang}, \bibinfo{person}{Xuesong He}, \bibinfo{person}{Xuyang Zhi},
  \bibinfo{person}{Hao Wang}, \bibinfo{person}{Xin Li},
  \bibinfo{person}{Feiyang Xu}, \bibinfo{person}{Deguang Liu},
  \bibinfo{person}{Huadong Liang}, \bibinfo{person}{Yi Li}, {et~al\mbox{.}}}
  \bibinfo{year}{2025}\natexlab{a}.
\newblock \showarticletitle{ChemEval: A Comprehensive Multi-Level Chemical
  Evaluation for Large Language Models}. In \bibinfo{booktitle}{\emph{The
  Thirteenth International Conference on Learning Representations}}.
\newblock
\urldef\tempurl%
\url{https://openreview.net/forum?id=JrqjSkEPrX}
\showURL{%
\tempurl}


\bibitem[Huang et~al\mbox{.}(2025b)]%
        {huang2025selfaug}
\bibfield{author}{\bibinfo{person}{Yuqing Huang}, \bibinfo{person}{Rongyang
  Zhang}, \bibinfo{person}{Qimeng Wang}, \bibinfo{person}{Chengqiang Lu},
  \bibinfo{person}{Yan Gao}, \bibinfo{person}{Yi Wu}, \bibinfo{person}{Yao Hu},
  \bibinfo{person}{Xuyang Zhi}, \bibinfo{person}{Guiquan Liu},
  \bibinfo{person}{Xin Li}, {et~al\mbox{.}}} \bibinfo{year}{2025}\natexlab{b}.
\newblock \showarticletitle{SelfAug: Mitigating Catastrophic Forgetting in
  Retrieval-Augmented Generation via Distribution Self-Alignment}. In
  \bibinfo{booktitle}{\emph{Findings of the Association for Computational
  Linguistics: EMNLP 2025}}. \bibinfo{pages}{14175--14190}.
\newblock
\href{https://doi.org/10.18653/v1/2025.findings-emnlp.763}{doi:\nolinkurl{10.18653/v1/2025.findings-emnlp.763}}


\bibitem[Ju et~al\mbox{.}(2025)]%
        {ju2025generative}
\bibfield{author}{\bibinfo{person}{Clark~Mingxuan Ju}, \bibinfo{person}{Liam
  Collins}, \bibinfo{person}{Leonardo Neves}, \bibinfo{person}{Bhuvesh Kumar},
  \bibinfo{person}{Louis~Yufeng Wang}, \bibinfo{person}{Tong Zhao}, {and}
  \bibinfo{person}{Neil Shah}.} \bibinfo{year}{2025}\natexlab{}.
\newblock \showarticletitle{Generative Recommendation with Semantic IDs: A
  Practitioner's Handbook}. In \bibinfo{booktitle}{\emph{Proceedings of the
  34th ACM International Conference on Information and Knowledge Management}}.
  \bibinfo{publisher}{Association for Computing Machinery},
  \bibinfo{address}{New York, NY, USA}, \bibinfo{pages}{6420--6425}.
\newblock


\bibitem[Kang and McAuley(2018)]%
        {kang2018self}
\bibfield{author}{\bibinfo{person}{Wang-Cheng Kang} {and}
  \bibinfo{person}{Julian McAuley}.} \bibinfo{year}{2018}\natexlab{}.
\newblock \showarticletitle{Self-attentive sequential recommendation}. In
  \bibinfo{booktitle}{\emph{2018 IEEE international conference on data mining
  (ICDM)}}. IEEE, \bibinfo{publisher}{IEEE}, \bibinfo{address}{Piscataway, NJ,
  USA}, \bibinfo{pages}{197--206}.
\newblock


\bibitem[Khrylchenko(2026)]%
        {khrylchenko2026variable}
\bibfield{author}{\bibinfo{person}{Kirill Khrylchenko}.}
  \bibinfo{year}{2026}\natexlab{}.
\newblock \showarticletitle{Variable-Length Semantic IDs for Recommender
  Systems}.
\newblock \bibinfo{journal}{\emph{arXiv preprint arXiv:2602.16375}}
  (\bibinfo{year}{2026}).
\newblock


\bibitem[Kim et~al\mbox{.}(2026)]%
        {kimtoken}
\bibfield{author}{\bibinfo{person}{Kibum Kim}, \bibinfo{person}{Sein Kim},
  \bibinfo{person}{HongSeok Kang}, \bibinfo{person}{Jiwan Kim},
  \bibinfo{person}{Heewoong Noh}, \bibinfo{person}{Yeonjun In},
  \bibinfo{person}{Kanghoon Yoon}, \bibinfo{person}{Jinoh Oh},
  \bibinfo{person}{Julian McAuley}, {and} \bibinfo{person}{Chanyoung Park}.}
  \bibinfo{year}{2026}\natexlab{}.
\newblock \showarticletitle{Token-Efficient Item Representation via Images for
  LLM Recommender Systems}. In \bibinfo{booktitle}{\emph{The Fourteenth
  International Conference on Learning Representations}}.
  \bibinfo{publisher}{International Conference on Learning Representations},
  \bibinfo{address}{Rio de Janeiro, Brazil}.
\newblock


\bibitem[Kudo(2018)]%
        {kudo2018subword}
\bibfield{author}{\bibinfo{person}{Taku Kudo}.}
  \bibinfo{year}{2018}\natexlab{}.
\newblock \showarticletitle{Subword regularization: Improving neural network
  translation models with multiple subword candidates}. In
  \bibinfo{booktitle}{\emph{Proceedings of the 56th Annual Meeting of the
  Association for Computational Linguistics (Volume 1: Long Papers)}}.
  \bibinfo{publisher}{Association for Computational Linguistics},
  \bibinfo{address}{Melbourne, Australia}, \bibinfo{pages}{66--75}.
\newblock


\bibitem[Kudo and Richardson(2018)]%
        {kudo2018sentencepiece}
\bibfield{author}{\bibinfo{person}{Taku Kudo} {and} \bibinfo{person}{John
  Richardson}.} \bibinfo{year}{2018}\natexlab{}.
\newblock \showarticletitle{SentencePiece: A simple and language independent
  subword tokenizer and detokenizer for neural text processing}. In
  \bibinfo{booktitle}{\emph{Proceedings of the 2018 conference on empirical
  methods in natural language processing: System demonstrations}}.
  \bibinfo{publisher}{Association for Computational Linguistics},
  \bibinfo{address}{Brussels, Belgium}, \bibinfo{pages}{66--71}.
\newblock


\bibitem[Lee et~al\mbox{.}(2022)]%
        {lee2022autoregressive}
\bibfield{author}{\bibinfo{person}{Doyup Lee}, \bibinfo{person}{Chiheon Kim},
  \bibinfo{person}{Saehoon Kim}, \bibinfo{person}{Minsu Cho}, {and}
  \bibinfo{person}{Wook-Shin Han}.} \bibinfo{year}{2022}\natexlab{}.
\newblock \showarticletitle{Autoregressive image generation using residual
  quantization}. In \bibinfo{booktitle}{\emph{Proceedings of the IEEE/CVF
  conference on computer vision and pattern recognition}}.
  \bibinfo{publisher}{IEEE/CVF}, \bibinfo{address}{New Orleans, LA, USA},
  \bibinfo{pages}{11523--11532}.
\newblock


\bibitem[Li et~al\mbox{.}(2026)]%
        {li2026unigrec}
\bibfield{author}{\bibinfo{person}{Jialei Li}, \bibinfo{person}{Yang Zhang},
  \bibinfo{person}{Yimeng Bai}, \bibinfo{person}{Shuai Zhu},
  \bibinfo{person}{Ziqi Xue}, \bibinfo{person}{Xiaoyan Zhao},
  \bibinfo{person}{Dingxian Wang}, \bibinfo{person}{Frank Yang},
  \bibinfo{person}{Andrew Rabinovich}, {and} \bibinfo{person}{Xiangnan He}.}
  \bibinfo{year}{2026}\natexlab{}.
\newblock \showarticletitle{UniGRec: Unified Generative Recommendation with
  Soft Identifiers for End-to-End Optimization}.
\newblock \bibinfo{journal}{\emph{arXiv preprint arXiv:2601.17438}}
  (\bibinfo{year}{2026}).
\newblock


\bibitem[Li et~al\mbox{.}(2025)]%
        {li2025survey}
\bibfield{author}{\bibinfo{person}{Xiaopeng Li}, \bibinfo{person}{Bo Chen},
  \bibinfo{person}{Junda She}, \bibinfo{person}{Shiteng Cao},
  \bibinfo{person}{You Wang}, \bibinfo{person}{Qinlin Jia},
  \bibinfo{person}{Haiying He}, \bibinfo{person}{Zheli Zhou},
  \bibinfo{person}{Zhao Liu}, \bibinfo{person}{Ji Liu}, {et~al\mbox{.}}}
  \bibinfo{year}{2025}\natexlab{}.
\newblock \showarticletitle{A survey of generative recommendation from a
  tri-decoupled perspective: Tokenization, architecture, and optimization}.
\newblock  (\bibinfo{year}{2025}).
\newblock
\href{https://doi.org/10.20944/preprints202512.0203.v1}{doi:\nolinkurl{10.20944/preprints202512.0203.v1}}


\bibitem[Liang et~al\mbox{.}(2025)]%
        {liang2025adaptive}
\bibfield{author}{\bibinfo{person}{Sheng Liang}, \bibinfo{person}{Hang Lv},
  \bibinfo{person}{Zhihao Wen}, \bibinfo{person}{Yaxiong Wu},
  \bibinfo{person}{Yongyue Zhang}, \bibinfo{person}{Hao Wang}, {and}
  \bibinfo{person}{Yong Li}.} \bibinfo{year}{2025}\natexlab{}.
\newblock \showarticletitle{Adaptive Schema-Aware Event Extraction with
  Retrieval-Augmented Generation}. In \bibinfo{booktitle}{\emph{Findings of the
  Association for Computational Linguistics: EMNLP 2025}}.
  \bibinfo{pages}{7927--7946}.
\newblock
\href{https://doi.org/10.18653/v1/2025.findings-emnlp.419}{doi:\nolinkurl{10.18653/v1/2025.findings-emnlp.419}}


\bibitem[Liu et~al\mbox{.}(2025)]%
        {liu2025generative}
\bibfield{author}{\bibinfo{person}{Enze Liu}, \bibinfo{person}{Bowen Zheng},
  \bibinfo{person}{Cheng Ling}, \bibinfo{person}{Lantao Hu},
  \bibinfo{person}{Han Li}, {and} \bibinfo{person}{Wayne~Xin Zhao}.}
  \bibinfo{year}{2025}\natexlab{}.
\newblock \showarticletitle{Generative recommender with end-to-end learnable
  item tokenization}. In \bibinfo{booktitle}{\emph{Proceedings of the 48th
  International ACM SIGIR Conference on Research and Development in Information
  Retrieval}}. \bibinfo{publisher}{Association for Computing Machinery},
  \bibinfo{address}{New York, NY, USA}, \bibinfo{pages}{729--739}.
\newblock


\bibitem[Lowerre(1976)]%
        {lowerre1976harpy}
\bibfield{author}{\bibinfo{person}{Bruce~T Lowerre}.}
  \bibinfo{year}{1976}\natexlab{}.
\newblock \bibinfo{booktitle}{\emph{The harpy speech recognition system.}}
\newblock \bibinfo{publisher}{Carnegie Mellon University},
  \bibinfo{address}{Pittsburgh, PA, USA}.
\newblock


\bibitem[Mielke et~al\mbox{.}(2021)]%
        {mielke2021between}
\bibfield{author}{\bibinfo{person}{Sabrina~J Mielke}, \bibinfo{person}{Zaid
  Alyafeai}, \bibinfo{person}{Elizabeth Salesky}, \bibinfo{person}{Colin
  Raffel}, \bibinfo{person}{Manan Dey}, \bibinfo{person}{Matthias Gall{\'e}},
  \bibinfo{person}{Arun Raja}, \bibinfo{person}{Chenglei Si},
  \bibinfo{person}{Wilson~Y Lee}, \bibinfo{person}{Beno{\^\i}t Sagot},
  {et~al\mbox{.}}} \bibinfo{year}{2021}\natexlab{}.
\newblock \showarticletitle{Between words and characters: A brief history of
  open-vocabulary modeling and tokenization in NLP}.
\newblock \bibinfo{journal}{\emph{arXiv preprint arXiv:2112.10508}}
  (\bibinfo{year}{2021}).
\newblock


\bibitem[Ni et~al\mbox{.}(2022)]%
        {ni2022sentence}
\bibfield{author}{\bibinfo{person}{Jianmo Ni},
  \bibinfo{person}{Gustavo~Hernandez Abrego}, \bibinfo{person}{Noah Constant},
  \bibinfo{person}{Ji Ma}, \bibinfo{person}{Keith Hall},
  \bibinfo{person}{Daniel Cer}, {and} \bibinfo{person}{Yinfei Yang}.}
  \bibinfo{year}{2022}\natexlab{}.
\newblock \showarticletitle{Sentence-t5: Scalable sentence encoders from
  pre-trained text-to-text models}. In \bibinfo{booktitle}{\emph{Findings of
  the association for computational linguistics: ACL 2022}}.
  \bibinfo{publisher}{Association for Computational Linguistics},
  \bibinfo{address}{Dublin, Ireland}, \bibinfo{pages}{1864--1874}.
\newblock


\bibitem[Pan et~al\mbox{.}(2025)]%
        {pan2025revisiting}
\bibfield{author}{\bibinfo{person}{Qiushi Pan}, \bibinfo{person}{Hao Wang},
  \bibinfo{person}{Guoyuan An}, \bibinfo{person}{Luankang Zhang},
  \bibinfo{person}{Wei Guo}, {and} \bibinfo{person}{Yong Liu}.}
  \bibinfo{year}{2025}\natexlab{}.
\newblock \showarticletitle{Revisiting Scalable Sequential Recommendation with
  Multi-Embedding Approach and Mixture-of-Experts}.
\newblock \bibinfo{journal}{\emph{arXiv preprint arXiv:2510.25285}}
  (\bibinfo{year}{2025}).
\newblock


\bibitem[Raffel et~al\mbox{.}(2020)]%
        {raffel2020exploring}
\bibfield{author}{\bibinfo{person}{Colin Raffel}, \bibinfo{person}{Noam
  Shazeer}, \bibinfo{person}{Adam Roberts}, \bibinfo{person}{Katherine Lee},
  \bibinfo{person}{Sharan Narang}, \bibinfo{person}{Michael Matena},
  \bibinfo{person}{Yanqi Zhou}, \bibinfo{person}{Wei Li}, {and}
  \bibinfo{person}{Peter~J Liu}.} \bibinfo{year}{2020}\natexlab{}.
\newblock \showarticletitle{Exploring the limits of transfer learning with a
  unified text-to-text transformer}.
\newblock \bibinfo{journal}{\emph{Journal of machine learning research}}
  \bibinfo{volume}{21}, \bibinfo{number}{140} (\bibinfo{year}{2020}),
  \bibinfo{pages}{1--67}.
\newblock


\bibitem[Rajaraman et~al\mbox{.}(2024)]%
        {rajaraman2024toward}
\bibfield{author}{\bibinfo{person}{Nived Rajaraman}, \bibinfo{person}{Jiantao
  Jiao}, {and} \bibinfo{person}{Kannan Ramchandran}.}
  \bibinfo{year}{2024}\natexlab{}.
\newblock \showarticletitle{Toward a theory of tokenization in llms}.
\newblock \bibinfo{journal}{\emph{arXiv preprint arXiv:2404.08335}}
  (\bibinfo{year}{2024}).
\newblock


\bibitem[Rajput et~al\mbox{.}(2023)]%
        {rajput2023recommender}
\bibfield{author}{\bibinfo{person}{Shashank Rajput}, \bibinfo{person}{Nikhil
  Mehta}, \bibinfo{person}{Anima Singh}, \bibinfo{person}{Raghunandan
  Hulikal~Keshavan}, \bibinfo{person}{Trung Vu}, \bibinfo{person}{Lukasz
  Heldt}, \bibinfo{person}{Lichan Hong}, \bibinfo{person}{Yi Tay},
  \bibinfo{person}{Vinh Tran}, \bibinfo{person}{Jonah Samost}, {et~al\mbox{.}}}
  \bibinfo{year}{2023}\natexlab{}.
\newblock \showarticletitle{Recommender systems with generative retrieval}.
\newblock \bibinfo{journal}{\emph{Advances in Neural Information Processing
  Systems}}  \bibinfo{volume}{36} (\bibinfo{year}{2023}),
  \bibinfo{pages}{10299--10315}.
\newblock


\bibitem[Sennrich et~al\mbox{.}(2016)]%
        {sennrich2016neural}
\bibfield{author}{\bibinfo{person}{Rico Sennrich}, \bibinfo{person}{Barry
  Haddow}, {and} \bibinfo{person}{Alexandra Birch}.}
  \bibinfo{year}{2016}\natexlab{}.
\newblock \showarticletitle{Neural machine translation of rare words with
  subword units}. In \bibinfo{booktitle}{\emph{Proceedings of the 54th annual
  meeting of the association for computational linguistics (volume 1: long
  papers)}}. \bibinfo{publisher}{Association for Computational Linguistics},
  \bibinfo{address}{Berlin, Germany}, \bibinfo{pages}{1715--1725}.
\newblock


\bibitem[Shen et~al\mbox{.}(2025)]%
        {shen2025plaw}
\bibfield{author}{\bibinfo{person}{Tingjia Shen}, \bibinfo{person}{Hao Wang},
  \bibinfo{person}{Chuhan Wu}, \bibinfo{person}{Jin~Yao Chin},
  \bibinfo{person}{Wei Guo}, \bibinfo{person}{Yong Liu},
  \bibinfo{person}{Huifeng Guo}, \bibinfo{person}{Defu Lian},
  \bibinfo{person}{Ruiming Tang}, {and} \bibinfo{person}{Enhong Chen}.}
  \bibinfo{year}{2025}\natexlab{}.
\newblock \showarticletitle{P-Law: Predicting Quantitative Scaling Law with
  Entropy Guidance in Large Recommendation Models}. In
  \bibinfo{booktitle}{\emph{Advances in Neural Information Processing
  Systems}}, Vol.~\bibinfo{volume}{38}. \bibinfo{pages}{14510--14539}.
\newblock
\href{https://doi.org/10.52202/085713-0436}{doi:\nolinkurl{10.52202/085713-0436}}


\bibitem[Sun et~al\mbox{.}(2019)]%
        {sun2019bert4rec}
\bibfield{author}{\bibinfo{person}{Fei Sun}, \bibinfo{person}{Jun Liu},
  \bibinfo{person}{Jian Wu}, \bibinfo{person}{Changhua Pei},
  \bibinfo{person}{Xiao Lin}, \bibinfo{person}{Wenwu Ou}, {and}
  \bibinfo{person}{Peng Jiang}.} \bibinfo{year}{2019}\natexlab{}.
\newblock \showarticletitle{BERT4Rec: Sequential recommendation with
  bidirectional encoder representations from transformer}. In
  \bibinfo{booktitle}{\emph{Proceedings of the 28th ACM international
  conference on information and knowledge management}}.
  \bibinfo{publisher}{Association for Computing Machinery},
  \bibinfo{address}{New York, NY, USA}, \bibinfo{pages}{1441--1450}.
\newblock


\bibitem[Sun et~al\mbox{.}(2023)]%
        {sun2023learning}
\bibfield{author}{\bibinfo{person}{Weiwei Sun}, \bibinfo{person}{Lingyong Yan},
  \bibinfo{person}{Zheng Chen}, \bibinfo{person}{Shuaiqiang Wang},
  \bibinfo{person}{Haichao Zhu}, \bibinfo{person}{Pengjie Ren},
  \bibinfo{person}{Zhumin Chen}, \bibinfo{person}{Dawei Yin},
  \bibinfo{person}{Maarten Rijke}, {and} \bibinfo{person}{Zhaochun Ren}.}
  \bibinfo{year}{2023}\natexlab{}.
\newblock \showarticletitle{Learning to tokenize for generative retrieval}.
\newblock \bibinfo{journal}{\emph{Advances in Neural Information Processing
  Systems}}  \bibinfo{volume}{36} (\bibinfo{year}{2023}),
  \bibinfo{pages}{46345--46361}.
\newblock


\bibitem[Tan et~al\mbox{.}(2024)]%
        {tan2024idgenrec}
\bibfield{author}{\bibinfo{person}{Juntao Tan}, \bibinfo{person}{Shuyuan Xu},
  \bibinfo{person}{Wenyue Hua}, \bibinfo{person}{Yingqiang Ge},
  \bibinfo{person}{Zelong Li}, {and} \bibinfo{person}{Yongfeng Zhang}.}
  \bibinfo{year}{2024}\natexlab{}.
\newblock \showarticletitle{Idgenrec: Llm-recsys alignment with textual id
  learning}. In \bibinfo{booktitle}{\emph{Proceedings of the 47th international
  ACM SIGIR conference on research and development in information retrieval}}.
  \bibinfo{publisher}{Association for Computing Machinery},
  \bibinfo{address}{New York, NY, USA}, \bibinfo{pages}{355--364}.
\newblock


\bibitem[Tang and Wang(2018)]%
        {tang2018personalized}
\bibfield{author}{\bibinfo{person}{Jiaxi Tang} {and} \bibinfo{person}{Ke
  Wang}.} \bibinfo{year}{2018}\natexlab{}.
\newblock \showarticletitle{Personalized top-n sequential recommendation via
  convolutional sequence embedding}. In \bibinfo{booktitle}{\emph{Proceedings
  of the eleventh ACM international conference on web search and data mining}}.
  \bibinfo{publisher}{Association for Computing Machinery},
  \bibinfo{address}{New York, NY, USA}, \bibinfo{pages}{565--573}.
\newblock


\bibitem[Tay et~al\mbox{.}(2022)]%
        {tay2022transformer}
\bibfield{author}{\bibinfo{person}{Yi Tay}, \bibinfo{person}{Vinh Tran},
  \bibinfo{person}{Mostafa Dehghani}, \bibinfo{person}{Jianmo Ni},
  \bibinfo{person}{Dara Bahri}, \bibinfo{person}{Harsh Mehta},
  \bibinfo{person}{Zhen Qin}, \bibinfo{person}{Kai Hui}, \bibinfo{person}{Zhe
  Zhao}, \bibinfo{person}{Jai Gupta}, {et~al\mbox{.}}}
  \bibinfo{year}{2022}\natexlab{}.
\newblock \showarticletitle{Transformer memory as a differentiable search
  index}.
\newblock \bibinfo{journal}{\emph{Advances in neural information processing
  systems}}  \bibinfo{volume}{35} (\bibinfo{year}{2022}),
  \bibinfo{pages}{21831--21843}.
\newblock


\bibitem[Van Den~Oord et~al\mbox{.}(2017)]%
        {van2017neural}
\bibfield{author}{\bibinfo{person}{Aaron Van Den~Oord}, \bibinfo{person}{Oriol
  Vinyals}, {et~al\mbox{.}}} \bibinfo{year}{2017}\natexlab{}.
\newblock \showarticletitle{Neural discrete representation learning}.
\newblock \bibinfo{journal}{\emph{Advances in neural information processing
  systems}}  \bibinfo{volume}{30} (\bibinfo{year}{2017}),
  \bibinfo{pages}{6306--6315}.
\newblock


\bibitem[Wang et~al\mbox{.}(2025b)]%
        {wang2025generative}
\bibfield{author}{\bibinfo{person}{Hao Wang}, \bibinfo{person}{Wei Guo},
  \bibinfo{person}{Luankang Zhang}, \bibinfo{person}{Jin~Yao Chin},
  \bibinfo{person}{Yufei Ye}, \bibinfo{person}{Huifeng Guo},
  \bibinfo{person}{Yong Liu}, \bibinfo{person}{Defu Lian},
  \bibinfo{person}{Ruiming Tang}, {and} \bibinfo{person}{Enhong Chen}.}
  \bibinfo{year}{2025}\natexlab{b}.
\newblock \showarticletitle{Generative Large Recommendation Models: Emerging
  Trends in LLMs for Recommendation}. In \bibinfo{booktitle}{\emph{Companion
  Proceedings of the ACM on Web Conference 2025}}. \bibinfo{pages}{49--52}.
\newblock
\href{https://doi.org/10.1145/3701716.3715865}{doi:\nolinkurl{10.1145/3701716.3715865}}


\bibitem[Wang et~al\mbox{.}(2024b)]%
        {wang2024denoising}
\bibfield{author}{\bibinfo{person}{Hao Wang}, \bibinfo{person}{Yongqiang Han},
  \bibinfo{person}{Kefan Wang}, \bibinfo{person}{Kai Cheng},
  \bibinfo{person}{Zhen Wang}, \bibinfo{person}{Wei Guo}, \bibinfo{person}{Yong
  Liu}, \bibinfo{person}{Defu Lian}, {and} \bibinfo{person}{Enhong Chen}.}
  \bibinfo{year}{2024}\natexlab{b}.
\newblock \showarticletitle{Denoising Pre-Training and Customized Prompt
  Learning for Efficient Multi-Behavior Sequential Recommendation}.
\newblock \bibinfo{journal}{\emph{arXiv preprint arXiv:2408.11372}}
  (\bibinfo{year}{2024}).
\newblock


\bibitem[Wang et~al\mbox{.}(2021)]%
        {wang2021hypersorec}
\bibfield{author}{\bibinfo{person}{Hao Wang}, \bibinfo{person}{Defu Lian},
  \bibinfo{person}{Hanghang Tong}, \bibinfo{person}{Qi Liu},
  \bibinfo{person}{Zhenya Huang}, {and} \bibinfo{person}{Enhong Chen}.}
  \bibinfo{year}{2021}\natexlab{}.
\newblock \showarticletitle{HyperSoRec: Exploiting Hyperbolic User and Item
  Representations with Multiple Aspects for Social-Aware Recommendation}.
\newblock \bibinfo{journal}{\emph{ACM Transactions on Information Systems}}
  \bibinfo{volume}{40}, \bibinfo{number}{2} (\bibinfo{year}{2021}),
  \bibinfo{pages}{1--28}.
\newblock
\href{https://doi.org/10.1145/3463913}{doi:\nolinkurl{10.1145/3463913}}


\bibitem[Wang et~al\mbox{.}(2019)]%
        {wang2019mcne}
\bibfield{author}{\bibinfo{person}{Hao Wang}, \bibinfo{person}{Tong Xu},
  \bibinfo{person}{Qi Liu}, \bibinfo{person}{Defu Lian},
  \bibinfo{person}{Enhong Chen}, \bibinfo{person}{Dongfang Du},
  \bibinfo{person}{Han Wu}, {and} \bibinfo{person}{Wen Su}.}
  \bibinfo{year}{2019}\natexlab{}.
\newblock \showarticletitle{MCNE: An End-to-End Framework for Learning Multiple
  Conditional Network Representations of Social Network}. In
  \bibinfo{booktitle}{\emph{Proceedings of the 25th ACM SIGKDD International
  Conference on Knowledge Discovery and Data Mining}}.
  \bibinfo{pages}{1064--1072}.
\newblock
\href{https://doi.org/10.1145/3292500.3330931}{doi:\nolinkurl{10.1145/3292500.3330931}}


\bibitem[Wang et~al\mbox{.}(2026b)]%
        {wang2026towards}
\bibfield{author}{\bibinfo{person}{Huimu Wang}, \bibinfo{person}{Xingzhi Yao},
  \bibinfo{person}{Yiming Qiu}, \bibinfo{person}{Qinghong Zhang},
  \bibinfo{person}{Haotian Wang}, \bibinfo{person}{Yufan Cui},
  \bibinfo{person}{Songlin Wang}, \bibinfo{person}{Sulong Xu}, {and}
  \bibinfo{person}{Mingming Li}.} \bibinfo{year}{2026}\natexlab{b}.
\newblock \showarticletitle{Towards Efficient and Generalizable Retrieval:
  Adaptive Semantic Quantization and Residual Knowledge Transfer}.
\newblock \bibinfo{journal}{\emph{arXiv preprint arXiv:2602.23978}}
  (\bibinfo{year}{2026}).
\newblock


\bibitem[Wang et~al\mbox{.}(2025f)]%
        {wang2025mfgslae}
\bibfield{author}{\bibinfo{person}{Hao Wang}, \bibinfo{person}{Mingjia Yin},
  \bibinfo{person}{Luankang Zhang}, \bibinfo{person}{Sirui Zhao}, {and}
  \bibinfo{person}{Enhong Chen}.} \bibinfo{year}{2025}\natexlab{f}.
\newblock \showarticletitle{MF-GSLAE: A Multi-Factor User Representation
  Pre-Training Framework for Dual-Target Cross-Domain Recommendation}.
\newblock \bibinfo{journal}{\emph{ACM Transactions on Information Systems}}
  \bibinfo{volume}{43}, \bibinfo{number}{2} (\bibinfo{year}{2025}),
  \bibinfo{pages}{1--28}.
\newblock
\href{https://doi.org/10.1145/3690382}{doi:\nolinkurl{10.1145/3690382}}


\bibitem[Wang et~al\mbox{.}(2025e)]%
        {wang2025decorrelated}
\bibfield{author}{\bibinfo{person}{Jiancheng Wang}, \bibinfo{person}{Mingjia
  Yin}, \bibinfo{person}{Hao Wang}, {and} \bibinfo{person}{Enhong Chen}.}
  \bibinfo{year}{2025}\natexlab{e}.
\newblock \showarticletitle{Enhancing CTR Prediction with De-Correlated Expert
  Networks}.
\newblock \bibinfo{journal}{\emph{arXiv preprint arXiv:2505.17925}}
  (\bibinfo{year}{2025}).
\newblock


\bibitem[Wang et~al\mbox{.}(2025d)]%
        {wang2025dlf}
\bibfield{author}{\bibinfo{person}{Kefan Wang}, \bibinfo{person}{Hao Wang},
  \bibinfo{person}{Wei Guo}, \bibinfo{person}{Yong Liu},
  \bibinfo{person}{Jianghao Lin}, \bibinfo{person}{Defu Lian}, {and}
  \bibinfo{person}{Enhong Chen}.} \bibinfo{year}{2025}\natexlab{d}.
\newblock \showarticletitle{DLF: Enhancing Explicit-Implicit Interaction via
  Dynamic Low-Order-Aware Fusion for CTR Prediction}. In
  \bibinfo{booktitle}{\emph{Proceedings of the 48th International ACM SIGIR
  Conference on Research and Development in Information Retrieval}}.
  \bibinfo{pages}{2213--2223}.
\newblock
\href{https://doi.org/10.1145/3726302.3729956}{doi:\nolinkurl{10.1145/3726302.3729956}}


\bibitem[Wang et~al\mbox{.}(2026a)]%
        {wang2026mec}
\bibfield{author}{\bibinfo{person}{Kefan Wang}, \bibinfo{person}{Hao Wang},
  \bibinfo{person}{Kenan Song}, \bibinfo{person}{Wei Guo}, \bibinfo{person}{Kai
  Cheng}, \bibinfo{person}{Zhi Li}, \bibinfo{person}{Yong Liu},
  \bibinfo{person}{Defu Lian}, {and} \bibinfo{person}{Enhong Chen}.}
  \bibinfo{year}{2026}\natexlab{a}.
\newblock \showarticletitle{A Universal Framework for Compressing Embeddings in
  CTR Prediction}. In \bibinfo{booktitle}{\emph{International Conference on
  Database Systems for Advanced Applications}}. Springer,
  \bibinfo{pages}{84--100}.
\newblock
\href{https://doi.org/10.1007/978-981-95-3830-0_6}{doi:\nolinkurl{10.1007/978-981-95-3830-0_6}}


\bibitem[Wang et~al\mbox{.}(2024a)]%
        {wang2024learnable}
\bibfield{author}{\bibinfo{person}{Wenjie Wang}, \bibinfo{person}{Honghui Bao},
  \bibinfo{person}{Xinyu Lin}, \bibinfo{person}{Jizhi Zhang},
  \bibinfo{person}{Yongqi Li}, \bibinfo{person}{Fuli Feng},
  \bibinfo{person}{See-Kiong Ng}, {and} \bibinfo{person}{Tat-Seng Chua}.}
  \bibinfo{year}{2024}\natexlab{a}.
\newblock \showarticletitle{Learnable item tokenization for generative
  recommendation}. In \bibinfo{booktitle}{\emph{Proceedings of the 33rd ACM
  International Conference on Information and Knowledge Management}}.
  \bibinfo{publisher}{Association for Computing Machinery},
  \bibinfo{address}{New York, NY, USA}, \bibinfo{pages}{2400--2409}.
\newblock


\bibitem[Wang et~al\mbox{.}(2025a)]%
        {wang2025act}
\bibfield{author}{\bibinfo{person}{Yifan Wang}, \bibinfo{person}{Weinan Gan},
  \bibinfo{person}{Longtao Xiao}, \bibinfo{person}{Jieming Zhu},
  \bibinfo{person}{Heng Chang}, \bibinfo{person}{Haozhao Wang},
  \bibinfo{person}{Rui Zhang}, \bibinfo{person}{Zhenhua Dong},
  \bibinfo{person}{Ruiming Tang}, {and} \bibinfo{person}{Ruixuan Li}.}
  \bibinfo{year}{2025}\natexlab{a}.
\newblock \showarticletitle{Act-With-Think: Chunk Auto-Regressive Modeling for
  Generative Recommendation}.
\newblock \bibinfo{journal}{\emph{arXiv preprint arXiv:2506.23643}}
  (\bibinfo{year}{2025}).
\newblock


\bibitem[Wang et~al\mbox{.}(2022)]%
        {wang2022neural}
\bibfield{author}{\bibinfo{person}{Yujing Wang}, \bibinfo{person}{Yingyan Hou},
  \bibinfo{person}{Haonan Wang}, \bibinfo{person}{Ziming Miao},
  \bibinfo{person}{Shibin Wu}, \bibinfo{person}{Qi Chen},
  \bibinfo{person}{Yuqing Xia}, \bibinfo{person}{Chengmin Chi},
  \bibinfo{person}{Guoshuai Zhao}, \bibinfo{person}{Zheng Liu},
  {et~al\mbox{.}}} \bibinfo{year}{2022}\natexlab{}.
\newblock \showarticletitle{A neural corpus indexer for document retrieval}.
\newblock \bibinfo{journal}{\emph{Advances in Neural Information Processing
  Systems}}  \bibinfo{volume}{35} (\bibinfo{year}{2022}),
  \bibinfo{pages}{25600--25614}.
\newblock


\bibitem[Wang et~al\mbox{.}(2025c)]%
        {wang2025empowering}
\bibfield{author}{\bibinfo{person}{Yuhao Wang}, \bibinfo{person}{Junwei Pan},
  \bibinfo{person}{Xinhang Li}, \bibinfo{person}{Maolin Wang},
  \bibinfo{person}{Yuan Wang}, \bibinfo{person}{Yue Liu},
  \bibinfo{person}{Dapeng Liu}, \bibinfo{person}{Jie Jiang}, {and}
  \bibinfo{person}{Xiangyu Zhao}.} \bibinfo{year}{2025}\natexlab{c}.
\newblock \showarticletitle{Empowering large language model for sequential
  recommendation via multimodal embeddings and semantic ids}. In
  \bibinfo{booktitle}{\emph{Proceedings of the 34th ACM International
  Conference on Information and Knowledge Management}}.
  \bibinfo{publisher}{Association for Computing Machinery},
  \bibinfo{address}{New York, NY, USA}, \bibinfo{pages}{3209--3219}.
\newblock


\bibitem[Wang et~al\mbox{.}(2024c)]%
        {wang2024eager}
\bibfield{author}{\bibinfo{person}{Ye Wang}, \bibinfo{person}{Jiahao Xun},
  \bibinfo{person}{Minjie Hong}, \bibinfo{person}{Jieming Zhu},
  \bibinfo{person}{Tao Jin}, \bibinfo{person}{Wang Lin},
  \bibinfo{person}{Haoyuan Li}, \bibinfo{person}{Linjun Li},
  \bibinfo{person}{Yan Xia}, \bibinfo{person}{Zhou Zhao}, {et~al\mbox{.}}}
  \bibinfo{year}{2024}\natexlab{c}.
\newblock \showarticletitle{Eager: Two-stream generative recommender with
  behavior-semantic collaboration}. In \bibinfo{booktitle}{\emph{Proceedings of
  the 30th ACM SIGKDD Conference on Knowledge Discovery and Data Mining}}.
  \bibinfo{publisher}{Association for Computing Machinery},
  \bibinfo{address}{New York, NY, USA}, \bibinfo{pages}{3245--3254}.
\newblock


\bibitem[Wei et~al\mbox{.}(2026)]%
        {wei2026oneloc}
\bibfield{author}{\bibinfo{person}{Zhipeng Wei}, \bibinfo{person}{Kuo Cai},
  \bibinfo{person}{Junda She}, \bibinfo{person}{Jie Chen},
  \bibinfo{person}{Minghao Chen}, \bibinfo{person}{Yang Zeng},
  \bibinfo{person}{Qiang Luo}, \bibinfo{person}{Wencong Zeng},
  \bibinfo{person}{Ruiming Tang}, \bibinfo{person}{Kun Gai}, {et~al\mbox{.}}}
  \bibinfo{year}{2026}\natexlab{}.
\newblock \showarticletitle{Oneloc: Geo-aware generative recommender systems
  for local life service}. In \bibinfo{booktitle}{\emph{Proceedings of the
  Nineteenth ACM International Conference on Web Search and Data Mining}}.
  \bibinfo{publisher}{Association for Computing Machinery},
  \bibinfo{address}{New York, NY, USA}, \bibinfo{pages}{735--744}.
\newblock


\bibitem[Wu et~al\mbox{.}(2024)]%
        {wu2024survey}
\bibfield{author}{\bibinfo{person}{Likang Wu}, \bibinfo{person}{Zhi Zheng},
  \bibinfo{person}{Zhaopeng Qiu}, \bibinfo{person}{Hao Wang},
  \bibinfo{person}{Hongchao Gu}, \bibinfo{person}{Tingjia Shen},
  \bibinfo{person}{Chuan Qin}, \bibinfo{person}{Chen Zhu},
  \bibinfo{person}{Hengshu Zhu}, \bibinfo{person}{Qi Liu}, {et~al\mbox{.}}}
  \bibinfo{year}{2024}\natexlab{}.
\newblock \showarticletitle{A Survey on Large Language Models for
  Recommendation}.
\newblock \bibinfo{journal}{\emph{World Wide Web}} \bibinfo{volume}{27},
  \bibinfo{number}{5} (\bibinfo{year}{2024}), \bibinfo{pages}{60}.
\newblock
\href{https://doi.org/10.1007/s11280-024-01291-2}{doi:\nolinkurl{10.1007/s11280-024-01291-2}}


\bibitem[Xiao et~al\mbox{.}(2025)]%
        {xiao2025unger}
\bibfield{author}{\bibinfo{person}{Longtao Xiao}, \bibinfo{person}{Haozhao
  Wang}, \bibinfo{person}{Cheng Wang}, \bibinfo{person}{Linfei Ji},
  \bibinfo{person}{Yifan Wang}, \bibinfo{person}{Jieming Zhu},
  \bibinfo{person}{Zhenhua Dong}, \bibinfo{person}{Rui Zhang}, {and}
  \bibinfo{person}{Ruixuan Li}.} \bibinfo{year}{2025}\natexlab{}.
\newblock \showarticletitle{Unger: Generative recommendation with a unified
  code via semantic and collaborative integration}.
\newblock \bibinfo{journal}{\emph{ACM Transactions on Information Systems}}
  \bibinfo{volume}{44}, \bibinfo{number}{2} (\bibinfo{year}{2025}),
  \bibinfo{pages}{1--31}.
\newblock


\bibitem[Xie et~al\mbox{.}(2025)]%
        {xie2025bottleneck}
\bibfield{author}{\bibinfo{person}{Wenjia Xie}, \bibinfo{person}{Hao Wang},
  \bibinfo{person}{Minghao Fang}, \bibinfo{person}{Ruize Yu},
  \bibinfo{person}{Wei Guo}, \bibinfo{person}{Yong Liu}, \bibinfo{person}{Defu
  Lian}, {and} \bibinfo{person}{Enhong Chen}.} \bibinfo{year}{2025}\natexlab{}.
\newblock \showarticletitle{Breaking the Bottleneck: User-Specific Optimization
  and Real-Time Inference Integration for Sequential Recommendation}. In
  \bibinfo{booktitle}{\emph{Proceedings of the 31st ACM SIGKDD Conference on
  Knowledge Discovery and Data Mining}}. \bibinfo{pages}{3333--3343}.
\newblock
\href{https://doi.org/10.1145/3711896.3736865}{doi:\nolinkurl{10.1145/3711896.3736865}}


\bibitem[Xie et~al\mbox{.}(2024)]%
        {xie2024fuzzymodeling}
\bibfield{author}{\bibinfo{person}{Wenjia Xie}, \bibinfo{person}{Hao Wang},
  \bibinfo{person}{Luankang Zhang}, \bibinfo{person}{Rui Zhou},
  \bibinfo{person}{Defu Lian}, {and} \bibinfo{person}{Enhong Chen}.}
  \bibinfo{year}{2024}\natexlab{}.
\newblock \showarticletitle{Breaking Determinism: Fuzzy Modeling of Sequential
  Recommendation Using Discrete State Space Diffusion Model}. In
  \bibinfo{booktitle}{\emph{Advances in Neural Information Processing
  Systems}}, Vol.~\bibinfo{volume}{37}. \bibinfo{pages}{22720--22744}.
\newblock
\href{https://doi.org/10.52202/079017-0715}{doi:\nolinkurl{10.52202/079017-0715}}


\bibitem[Xu et~al\mbox{.}(2025)]%
        {xu2025mirrn}
\bibfield{author}{\bibinfo{person}{Xiang Xu}, \bibinfo{person}{Hao Wang},
  \bibinfo{person}{Wei Guo}, \bibinfo{person}{Luankang Zhang},
  \bibinfo{person}{Wanshan Yang}, \bibinfo{person}{Runlong Yu},
  \bibinfo{person}{Yong Liu}, \bibinfo{person}{Defu Lian}, {and}
  \bibinfo{person}{Enhong Chen}.} \bibinfo{year}{2025}\natexlab{}.
\newblock \showarticletitle{Multi-Granularity Interest Retrieval and Refinement
  Network for Long-Term User Behavior Modeling in CTR Prediction}. In
  \bibinfo{booktitle}{\emph{Proceedings of the 31st ACM SIGKDD Conference on
  Knowledge Discovery and Data Mining}}. \bibinfo{pages}{2745--2755}.
\newblock
\href{https://doi.org/10.1145/3690624.3709438}{doi:\nolinkurl{10.1145/3690624.3709438}}


\bibitem[Xu et~al\mbox{.}(2026)]%
        {xu2026mmq}
\bibfield{author}{\bibinfo{person}{Yi Xu}, \bibinfo{person}{Moyu Zhang},
  \bibinfo{person}{Chenxuan Li}, \bibinfo{person}{Zhihao Liao},
  \bibinfo{person}{Haibo Xing}, \bibinfo{person}{Hao Deng},
  \bibinfo{person}{Jinxin Hu}, \bibinfo{person}{Yu Zhang},
  \bibinfo{person}{Xiaoyi Zeng}, {and} \bibinfo{person}{Jing Zhang}.}
  \bibinfo{year}{2026}\natexlab{}.
\newblock \showarticletitle{Mmq: Multimodal mixture-of-quantization
  tokenization for semantic id generation and user behavioral adaptation}. In
  \bibinfo{booktitle}{\emph{Proceedings of the Nineteenth ACM International
  Conference on Web Search and Data Mining}}. \bibinfo{publisher}{Association
  for Computing Machinery}, \bibinfo{address}{New York, NY, USA},
  \bibinfo{pages}{788--797}.
\newblock


\bibitem[Yang et~al\mbox{.}(2026)]%
        {yang2026sparse}
\bibfield{author}{\bibinfo{person}{Yuhao Yang}, \bibinfo{person}{Zhi Ji},
  \bibinfo{person}{Zhaopeng Li}, \bibinfo{person}{Yi Li},
  \bibinfo{person}{Zhonglin Mo}, \bibinfo{person}{Yue Ding},
  \bibinfo{person}{Kai Chen}, \bibinfo{person}{Zijian Zhang},
  \bibinfo{person}{Jie Li}, \bibinfo{person}{LIU LIN}, {et~al\mbox{.}}}
  \bibinfo{year}{2026}\natexlab{}.
\newblock \showarticletitle{Sparse meets dense: Unified generative
  recommendations with cascaded sparse-dense representations}.
\newblock \bibinfo{journal}{\emph{Advances in Neural Information Processing
  Systems}}  \bibinfo{volume}{38} (\bibinfo{year}{2026}),
  \bibinfo{pages}{93746--93770}.
\newblock


\bibitem[Ye et~al\mbox{.}(2025a)]%
        {ye2025fuxialpha}
\bibfield{author}{\bibinfo{person}{Yufei Ye}, \bibinfo{person}{Wei Guo},
  \bibinfo{person}{Jin~Yao Chin}, \bibinfo{person}{Hao Wang},
  \bibinfo{person}{Hong Zhu}, \bibinfo{person}{Xi Lin}, \bibinfo{person}{Yuyang
  Ye}, \bibinfo{person}{Yong Liu}, \bibinfo{person}{Ruiming Tang},
  \bibinfo{person}{Defu Lian}, {et~al\mbox{.}}}
  \bibinfo{year}{2025}\natexlab{a}.
\newblock \showarticletitle{FuXi-$\\alpha$: Scaling Recommendation Model with
  Feature Interaction Enhanced Transformer}. In
  \bibinfo{booktitle}{\emph{Companion Proceedings of the ACM on Web Conference
  2025}}. \bibinfo{pages}{557--566}.
\newblock
\href{https://doi.org/10.1145/3701716.3715448}{doi:\nolinkurl{10.1145/3701716.3715448}}


\bibitem[Ye et~al\mbox{.}(2026)]%
        {ye2026fuxilinear}
\bibfield{author}{\bibinfo{person}{Yufei Ye}, \bibinfo{person}{Wei Guo},
  \bibinfo{person}{Hao Wang}, \bibinfo{person}{Luankang Zhang},
  \bibinfo{person}{Heng Chang}, \bibinfo{person}{Hong Zhu},
  \bibinfo{person}{Yuyang Ye}, \bibinfo{person}{Yong Liu},
  \bibinfo{person}{Defu Lian}, {and} \bibinfo{person}{Enhong Chen}.}
  \bibinfo{year}{2026}\natexlab{}.
\newblock \showarticletitle{Fuxi-linear: Unleashing the Power of Linear
  Attention in Long-Term Time-Aware Sequential Recommendation}. In
  \bibinfo{booktitle}{\emph{Proceedings of the 32nd ACM SIGKDD Conference on
  Knowledge Discovery and Data Mining}}. \bibinfo{pages}{6200--6210}.
\newblock
\href{https://doi.org/10.1145/3770855.3818139}{doi:\nolinkurl{10.1145/3770855.3818139}}


\bibitem[Ye et~al\mbox{.}(2025b)]%
        {ye2025fuxibeta}
\bibfield{author}{\bibinfo{person}{Yufei Ye}, \bibinfo{person}{Wei Guo},
  \bibinfo{person}{Hao Wang}, \bibinfo{person}{Hong Zhu},
  \bibinfo{person}{Yuyang Ye}, \bibinfo{person}{Yong Liu},
  \bibinfo{person}{Huifeng Guo}, \bibinfo{person}{Ruiming Tang},
  \bibinfo{person}{Defu Lian}, {and} \bibinfo{person}{Enhong Chen}.}
  \bibinfo{year}{2025}\natexlab{b}.
\newblock \showarticletitle{FuXi-$\\beta$: Towards a Lightweight and Fast
  Large-Scale Generative Recommendation Model}.
\newblock \bibinfo{journal}{\emph{arXiv preprint arXiv:2508.10615}}
  (\bibinfo{year}{2025}).
\newblock


\bibitem[Yin et~al\mbox{.}(2025)]%
        {yin2025genctr}
\bibfield{author}{\bibinfo{person}{Mingjia Yin}, \bibinfo{person}{Junwei Pan},
  \bibinfo{person}{Hao Wang}, \bibinfo{person}{Ximei Wang},
  \bibinfo{person}{Shangyu Zhang}, \bibinfo{person}{Jie Jiang},
  \bibinfo{person}{Defu Lian}, {and} \bibinfo{person}{Enhong Chen}.}
  \bibinfo{year}{2025}\natexlab{}.
\newblock \showarticletitle{From Feature Interaction to Feature Generation: A
  Generative Paradigm of CTR Prediction Models}. In
  \bibinfo{booktitle}{\emph{Proceedings of the 42nd International Conference on
  Machine Learning}} \emph{(\bibinfo{series}{Proceedings of Machine Learning
  Research}, Vol.~\bibinfo{volume}{267})}. \bibinfo{publisher}{PMLR},
  \bibinfo{pages}{72393--72413}.
\newblock
\urldef\tempurl%
\url{https://proceedings.mlr.press/v267/yin25c.html}
\showURL{%
\tempurl}


\bibitem[Yin et~al\mbox{.}(2024)]%
        {yin2024dr4sr}
\bibfield{author}{\bibinfo{person}{Mingjia Yin}, \bibinfo{person}{Hao Wang},
  \bibinfo{person}{Wei Guo}, \bibinfo{person}{Yong Liu},
  \bibinfo{person}{Suojuan Zhang}, \bibinfo{person}{Sirui Zhao},
  \bibinfo{person}{Defu Lian}, {and} \bibinfo{person}{Enhong Chen}.}
  \bibinfo{year}{2024}\natexlab{}.
\newblock \showarticletitle{Dataset Regeneration for Sequential
  Recommendation}. In \bibinfo{booktitle}{\emph{Proceedings of the 30th ACM
  SIGKDD Conference on Knowledge Discovery and Data Mining}}.
  \bibinfo{pages}{3954--3965}.
\newblock
\href{https://doi.org/10.1145/3637528.3671841}{doi:\nolinkurl{10.1145/3637528.3671841}}


\bibitem[Yu et~al\mbox{.}(2026)]%
        {yu2026planagent}
\bibfield{author}{\bibinfo{person}{Haocheng Yu}, \bibinfo{person}{Yaxiong Wu},
  \bibinfo{person}{Hao Wang}, \bibinfo{person}{Wei Guo}, \bibinfo{person}{Yong
  Liu}, \bibinfo{person}{Yawen Li}, \bibinfo{person}{Yuyang Ye},
  \bibinfo{person}{Junping Du}, {and} \bibinfo{person}{Enhong Chen}.}
  \bibinfo{year}{2026}\natexlab{}.
\newblock \showarticletitle{Thought-Augmented Planning for LLM-Powered
  Interactive Recommender Agent}. In \bibinfo{booktitle}{\emph{Proceedings of
  the 32nd ACM SIGKDD Conference on Knowledge Discovery and Data Mining}}.
  \bibinfo{pages}{1821--1832}.
\newblock
\href{https://doi.org/10.1145/3770854.3780286}{doi:\nolinkurl{10.1145/3770854.3780286}}


\bibitem[Zeghidour et~al\mbox{.}(2021)]%
        {zeghidour2021soundstream}
\bibfield{author}{\bibinfo{person}{Neil Zeghidour}, \bibinfo{person}{Alejandro
  Luebs}, \bibinfo{person}{Ahmed Omran}, \bibinfo{person}{Jan Skoglund}, {and}
  \bibinfo{person}{Marco Tagliasacchi}.} \bibinfo{year}{2021}\natexlab{}.
\newblock \showarticletitle{Soundstream: An end-to-end neural audio codec}.
\newblock \bibinfo{journal}{\emph{IEEE/ACM Transactions on Audio, Speech, and
  Language Processing}}  \bibinfo{volume}{30} (\bibinfo{year}{2021}),
  \bibinfo{pages}{495--507}.
\newblock


\bibitem[Zhang et~al\mbox{.}(2025b)]%
        {zhang2025gpr}
\bibfield{author}{\bibinfo{person}{Jun Zhang}, \bibinfo{person}{Yi Li},
  \bibinfo{person}{Yue Liu}, \bibinfo{person}{Changping Wang},
  \bibinfo{person}{Yuan Wang}, \bibinfo{person}{Yuling Xiong},
  \bibinfo{person}{Xun Liu}, \bibinfo{person}{Haiyang Wu},
  \bibinfo{person}{Qian Li}, \bibinfo{person}{Enming Zhang}, {et~al\mbox{.}}}
  \bibinfo{year}{2025}\natexlab{b}.
\newblock \showarticletitle{GPR: Towards a Generative Pre-trained One-Model
  Paradigm for Large-Scale Advertising Recommendation}.
\newblock \bibinfo{journal}{\emph{arXiv preprint arXiv:2511.10138}}
  (\bibinfo{year}{2025}).
\newblock


\bibitem[Zhang et~al\mbox{.}(2025f)]%
        {zhang2025td3}
\bibfield{author}{\bibinfo{person}{Jiaqing Zhang}, \bibinfo{person}{Mingjia
  Yin}, \bibinfo{person}{Hao Wang}, \bibinfo{person}{Yawen Li},
  \bibinfo{person}{Yuyang Ye}, \bibinfo{person}{Xingyu Lou},
  \bibinfo{person}{Junping Du}, {and} \bibinfo{person}{Enhong Chen}.}
  \bibinfo{year}{2025}\natexlab{f}.
\newblock \showarticletitle{TD3: Tucker Decomposition Based Dataset
  Distillation Method for Sequential Recommendation}. In
  \bibinfo{booktitle}{\emph{Companion Proceedings of the ACM on Web Conference
  2025}}. \bibinfo{pages}{3994--4003}.
\newblock
\href{https://doi.org/10.1145/3696410.3714613}{doi:\nolinkurl{10.1145/3696410.3714613}}


\bibitem[Zhang et~al\mbox{.}(2026a)]%
        {zhang2026whythinking}
\bibfield{author}{\bibinfo{person}{Luankang Zhang}, \bibinfo{person}{Yonghao
  Huang}, \bibinfo{person}{Hang Lv}, \bibinfo{person}{Xuyang Zhi},
  \bibinfo{person}{Mingjia Yin}, \bibinfo{person}{Yuyang Ye},
  \bibinfo{person}{Wei Guo}, \bibinfo{person}{Hao Wang}, {and}
  \bibinfo{person}{Enhong Chen}.} \bibinfo{year}{2026}\natexlab{a}.
\newblock \showarticletitle{Why Thinking Hurts: Diagnosing and Rectifying
  Linguistic Inertia in Large Language Models for Recommendation}.
\newblock \bibinfo{journal}{\emph{arXiv preprint arXiv:2602.16587}}
  (\bibinfo{year}{2026}).
\newblock


\bibitem[Zhang et~al\mbox{.}(2026b)]%
        {zhang2026useragent}
\bibfield{author}{\bibinfo{person}{Luankang Zhang}, \bibinfo{person}{Hang Lv},
  \bibinfo{person}{Qiushi Pan}, \bibinfo{person}{Kefen Wang},
  \bibinfo{person}{Yonghao Huang}, \bibinfo{person}{Xinrui Miao},
  \bibinfo{person}{Yin Xu}, \bibinfo{person}{Wei Guo}, \bibinfo{person}{Yong
  Liu}, \bibinfo{person}{Hao Wang}, {et~al\mbox{.}}}
  \bibinfo{year}{2026}\natexlab{b}.
\newblock \showarticletitle{The Next Paradigm Is User-Centric Agent, Not
  Platform-Centric Service}.
\newblock \bibinfo{journal}{\emph{arXiv preprint arXiv:2602.15682}}
  (\bibinfo{year}{2026}).
\newblock


\bibitem[Zhang et~al\mbox{.}(2025d)]%
        {zhang2025killing}
\bibfield{author}{\bibinfo{person}{Luankang Zhang}, \bibinfo{person}{Kenan
  Song}, \bibinfo{person}{Yi~Quan Lee}, \bibinfo{person}{Wei Guo},
  \bibinfo{person}{Hao Wang}, \bibinfo{person}{Yawen Li},
  \bibinfo{person}{Huifeng Guo}, \bibinfo{person}{Yong Liu},
  \bibinfo{person}{Defu Lian}, {and} \bibinfo{person}{Enhong Chen}.}
  \bibinfo{year}{2025}\natexlab{d}.
\newblock \showarticletitle{Killing Two Birds with One Stone: Unifying
  Retrieval and Ranking with a Single Generative Recommendation Model}. In
  \bibinfo{booktitle}{\emph{Proceedings of the 48th International ACM SIGIR
  Conference on Research and Development in Information Retrieval}}.
  \bibinfo{pages}{2224--2234}.
\newblock
\href{https://doi.org/10.1145/3726302.3730017}{doi:\nolinkurl{10.1145/3726302.3730017}}


\bibitem[Zhang et~al\mbox{.}(2026c)]%
        {zhang2026can}
\bibfield{author}{\bibinfo{person}{Luankang Zhang}, \bibinfo{person}{Hao Wang},
  \bibinfo{person}{Zhongzhou Liu}, \bibinfo{person}{Mingjia Yin},
  \bibinfo{person}{Yonghao Huang}, \bibinfo{person}{Jiaqi Li},
  \bibinfo{person}{Wei Guo}, \bibinfo{person}{Yong Liu},
  \bibinfo{person}{Huifeng Guo}, \bibinfo{person}{Defu Lian}, {et~al\mbox{.}}}
  \bibinfo{year}{2026}\natexlab{c}.
\newblock \showarticletitle{Can Recommender Systems Teach Themselves? A
  Recursive Self-Improving Framework with Fidelity Control}.
\newblock \bibinfo{journal}{\emph{arXiv preprint arXiv:2602.15659}}
  (\bibinfo{year}{2026}).
\newblock
\shownote{Accepted to ICML 2026}.
\newblock
\urldef\tempurl%
\url{https://arxiv.org/abs/2602.15659}
\showURL{%
\tempurl}


\bibitem[Zhang et~al\mbox{.}(2025e)]%
        {zhang2025unified}
\bibfield{author}{\bibinfo{person}{Luankang Zhang}, \bibinfo{person}{Hao Wang},
  \bibinfo{person}{Suojuan Zhang}, \bibinfo{person}{Mingjia Yin},
  \bibinfo{person}{Yongqiang Han}, \bibinfo{person}{Jiaqing Zhang},
  \bibinfo{person}{Defu Lian}, {and} \bibinfo{person}{Enhong Chen}.}
  \bibinfo{year}{2025}\natexlab{e}.
\newblock \showarticletitle{A Unified Framework for Adaptive Representation
  Enhancement and Inversed Learning in Cross-Domain Recommendation}. In
  \bibinfo{booktitle}{\emph{International Conference on Database Systems for
  Advanced Applications}}. Springer, \bibinfo{pages}{115--130}.
\newblock
\href{https://doi.org/10.1007/978-981-97-5555-4_8}{doi:\nolinkurl{10.1007/978-981-97-5555-4_8}}


\bibitem[Zhang et~al\mbox{.}(2025a)]%
        {zhang2025ragigbench}
\bibfield{author}{\bibinfo{person}{Rongyang Zhang}, \bibinfo{person}{Yuqing
  Huang}, \bibinfo{person}{Chengqiang Lu}, \bibinfo{person}{Qimeng Wang},
  \bibinfo{person}{Yan Gao}, \bibinfo{person}{Yao Hu}, \bibinfo{person}{Yin
  Xu}, \bibinfo{person}{Wei Wang}, \bibinfo{person}{Hao Wang}, {and}
  \bibinfo{person}{Enhong Chen}.} \bibinfo{year}{2025}\natexlab{a}.
\newblock \showarticletitle{RAG-IGBench: Innovative Evaluation for RAG-Based
  Interleaved Generation in Open-Domain Question Answering}. In
  \bibinfo{booktitle}{\emph{Advances in Neural Information Processing
  Systems}}, Vol.~\bibinfo{volume}{38}. \bibinfo{pages}{135329--135355}.
\newblock
\href{https://doi.org/10.52202/085713-4071}{doi:\nolinkurl{10.52202/085713-4071}}


\bibitem[Zhang et~al\mbox{.}(2025c)]%
        {zhang2025purely}
\bibfield{author}{\bibinfo{person}{Ruohan Zhang}, \bibinfo{person}{Jiacheng
  Li}, \bibinfo{person}{Julian McAuley}, {and} \bibinfo{person}{Yupeng Hou}.}
  \bibinfo{year}{2025}\natexlab{c}.
\newblock \showarticletitle{Purely Semantic Indexing for LLM-based Generative
  Recommendation and Retrieval}.
\newblock \bibinfo{journal}{\emph{arXiv preprint arXiv:2509.16446}}
  (\bibinfo{year}{2025}).
\newblock


\bibitem[Zhao et~al\mbox{.}(2023)]%
        {zhao2023survey}
\bibfield{author}{\bibinfo{person}{Wayne~Xin Zhao}, \bibinfo{person}{Kun Zhou},
  \bibinfo{person}{Junyi Li}, \bibinfo{person}{Tianyi Tang},
  \bibinfo{person}{Xiaolei Wang}, \bibinfo{person}{Yupeng Hou},
  \bibinfo{person}{Yingqian Min}, \bibinfo{person}{Beichen Zhang},
  \bibinfo{person}{Junjie Zhang}, \bibinfo{person}{Zican Dong},
  {et~al\mbox{.}}} \bibinfo{year}{2023}\natexlab{}.
\newblock \showarticletitle{A survey of large language models}.
\newblock \bibinfo{journal}{\emph{arXiv preprint arXiv:2303.18223}}
  \bibinfo{volume}{1}, \bibinfo{number}{2} (\bibinfo{year}{2023}),
  \bibinfo{pages}{1--124}.
\newblock


\bibitem[Zheng et~al\mbox{.}(2024)]%
        {zheng2024adapting}
\bibfield{author}{\bibinfo{person}{Bowen Zheng}, \bibinfo{person}{Yupeng Hou},
  \bibinfo{person}{Hongyu Lu}, \bibinfo{person}{Yu Chen},
  \bibinfo{person}{Wayne~Xin Zhao}, \bibinfo{person}{Ming Chen}, {and}
  \bibinfo{person}{Ji-Rong Wen}.} \bibinfo{year}{2024}\natexlab{}.
\newblock \showarticletitle{Adapting large language models by integrating
  collaborative semantics for recommendation}. In
  \bibinfo{booktitle}{\emph{2024 IEEE 40th International Conference on Data
  Engineering (ICDE)}}. IEEE, \bibinfo{publisher}{IEEE},
  \bibinfo{address}{Piscataway, NJ, USA}, \bibinfo{pages}{1435--1448}.
\newblock


\bibitem[Zhou et~al\mbox{.}(2025a)]%
        {zhou2025onerec}
\bibfield{author}{\bibinfo{person}{Guorui Zhou}, \bibinfo{person}{Jiaxin Deng},
  \bibinfo{person}{Jinghao Zhang}, \bibinfo{person}{Kuo Cai},
  \bibinfo{person}{Lejian Ren}, \bibinfo{person}{Qiang Luo},
  \bibinfo{person}{Qianqian Wang}, \bibinfo{person}{Qigen Hu},
  \bibinfo{person}{Rui Huang}, \bibinfo{person}{Shiyao Wang}, {et~al\mbox{.}}}
  \bibinfo{year}{2025}\natexlab{a}.
\newblock \showarticletitle{Onerec technical report}.
\newblock \bibinfo{journal}{\emph{arXiv preprint arXiv:2506.13695}}
  (\bibinfo{year}{2025}).
\newblock


\bibitem[Zhou et~al\mbox{.}(2025b)]%
        {zhou2025onerec2}
\bibfield{author}{\bibinfo{person}{Guorui Zhou}, \bibinfo{person}{Hengrui Hu},
  \bibinfo{person}{Hongtao Cheng}, \bibinfo{person}{Huanjie Wang},
  \bibinfo{person}{Jiaxin Deng}, \bibinfo{person}{Jinghao Zhang},
  \bibinfo{person}{Kuo Cai}, \bibinfo{person}{Lejian Ren}, \bibinfo{person}{Lu
  Ren}, \bibinfo{person}{Liao Yu}, {et~al\mbox{.}}}
  \bibinfo{year}{2025}\natexlab{b}.
\newblock \showarticletitle{Onerec-v2 technical report}.
\newblock \bibinfo{journal}{\emph{arXiv preprint arXiv:2508.20900}}
  (\bibinfo{year}{2025}).
\newblock


\bibitem[Zhou et~al\mbox{.}(2026)]%
        {zhou2026survey}
\bibfield{author}{\bibinfo{person}{Rui Zhou}, \bibinfo{person}{Qinglin Jia},
  \bibinfo{person}{Bo Chen}, \bibinfo{person}{Peng Xu}, \bibinfo{person}{Yijia
  Sun}, \bibinfo{person}{Siyuan Lou}, \bibinfo{person}{Chaoxin Fu},
  \bibinfo{person}{Mengyuan Fu}, \bibinfo{person}{Guoming Shen},
  \bibinfo{person}{Zheli Zhou}, {et~al\mbox{.}}}
  \bibinfo{year}{2026}\natexlab{}.
\newblock \showarticletitle{A Survey of User Lifelong Behavior Modeling:
  Perspectives on Efficiency and Effectiveness}.
\newblock \bibinfo{journal}{\emph{Preprints.org}} (\bibinfo{date}{jan}
  \bibinfo{year}{2026}).
\newblock
\href{https://doi.org/10.20944/preprints202601.1559.v1}{doi:\nolinkurl{10.20944/preprints202601.1559.v1}}


\bibitem[Zhou et~al\mbox{.}(2025c)]%
        {zhou2025mit}
\bibfield{author}{\bibinfo{person}{Rui Zhou}, \bibinfo{person}{Hao Wang},
  \bibinfo{person}{Wei Guo}, \bibinfo{person}{Qinglin Jia},
  \bibinfo{person}{Wenjia Xie}, \bibinfo{person}{Xiang Xu},
  \bibinfo{person}{Yong Liu}, \bibinfo{person}{Defu Lian}, {and}
  \bibinfo{person}{Enhong Chen}.} \bibinfo{year}{2025}\natexlab{c}.
\newblock \showarticletitle{MIT: A Multi-Tower Information Transfer Framework
  Based on Hierarchical Task Relationship Modeling}. In
  \bibinfo{booktitle}{\emph{Companion Proceedings of the ACM on Web Conference
  2025}}. \bibinfo{pages}{651--660}.
\newblock
\href{https://doi.org/10.1145/3701716.3715249}{doi:\nolinkurl{10.1145/3701716.3715249}}


\end{thebibliography}

\end{document}